\documentclass[draft,numberedheadings]{aipproc}
\layoutstyle{8x11single}
\usepackage{graphicx}
\def\la{\mathrel{\mathpalette\fun <}}
\def\ga{\mathrel{\mathpalette\fun >}}
\def\fun#1#2{\lower3.6pt\vbox{\baselineskip0pt\lineskip.9pt
  \ialign{$\mathsurround=0pt#1\hfil##\hfil$\crcr#2\crcr\sim\crcr}}}
\def\ss{{\hbox{\boldmath$\sigma$}}}

\begin{document}
\title{Lectures on Astroparticle Physics}

\author{G\"unter Sigl}
{address={GReCO, Institut d'Astrophysique de Paris, C.N.R.S.\\
98 bis boulevard Arago, F-75014 Paris, France\\
F\'{e}d\'{e}ration de Recherche Astroparticule et Cosmologie, 
Universit\'{e} Paris 7\\
2 place Jussieu, 75251 Paris Cedex 05, France}}

\begin{abstract}
These are extended notes of a series of lectures given at the XIth
Brazilian School of Cosmology and Gravitation. They provide a selection
of topics at the intersection of particle and astrophysics. The first part
gives a short introduction to the theory of electroweak interactions,
with specific emphasize on neutrinos. In the second part we apply
this framework to selected topics in astrophysics and cosmology,
namely neutrino oscillations, neutrino hot dark dark matter, and big bang
nucleosynthesis. The last part is devoted to ultra high energy
cosmic rays and neutrinos where again particle physics aspects
are emphasized. The often complementary role of laboratory experiments
is also discussed in several examples.
\end{abstract}

\maketitle

\newpage
\tableofcontents
\newpage

\section{Introduction and Reminder: Fermi Theory of Weak Interactions}
Good introductory texts on particle physics are contained in
Ref.~\cite{perkins} (more phenomenologically and experimentally
oriented) and in Refs.~\cite{weinberg1,weinberg2,weinberg3}.
Here we will only recall the most essential facts.

We will usually use natural units in which $\hbar=c=k=1$, unless
these constants are explicitly given.

Neutrinos only have weak interactions. Historically, experiments
with neutrinos obtained from decaying pions and kaons have shown
that charged and neutral leptons appear in three doublets:

\begin{table}[ht]
\caption{The lepton doublets}\label{tab1}
\begin{tabular}{cccc}
q & $L_e=1$ & $L_\mu=1$ & $L_\tau=1$ \\
\hline \\
$\matrix{0\cr -1}$ & $\left(\matrix{\nu_e\cr e^-}\right)$ &
$\left(\matrix{\nu_\mu\cr \mu^-}\right)$ &
$\left(\matrix{\nu_\tau\cr \tau^-}\right)$ \\
\end{tabular}
\end{table}
Charge $q$ and lepton numbers $L_e$, $L_\mu$, and $L_\tau$ are
conserved separately. There are corresponding doublets
of anti-leptons with opposite charge and lepton numbers, denoted
by $\bar{\nu}_i$ for the anti-neutrinos and by the respective
positively charged anti-leptons.

Therefore, allowed reactions include $n\to p e^-\bar{\nu}_e$ (nuclear
$\beta$-decay), $\bar{\nu}_e p\to n e^+$ (inverse neutron decay)
$\pi^+\to\mu^+\nu_\mu$, $\nu_\mu n\to p\mu^-$, but exclude
$\nu_\mu p\to n\mu^+$, $\mu^+\to e^+\gamma$.

The ``neutrino'' is thus defined as the neutral particle emitted
together with positrons in $\beta^+$-decay or following K-capture
of electrons. The ``anti-neutrino'' accompanies negative electrons in
$\beta^-$-decay.

Lifetimes for weak decays are long compared to lifetimes associated
with electromagnetic ($\sim10^{-19}\,$s) and strong ($\sim10^{-23}\,$s)
interactions. A weak interaction cross section at $\sim1\,$GeV
interaction energy is typically $\sim10^{12}$ times smaller than
a strong interaction cross section.

Weak interactions are classified into leptonic, semi-leptonic,
and non-leptonic interactions.

{\it Fermi's golden rule} yields for the rate $\Gamma$ of a reaction
from an initial state $i$ to a final state $f$ the expression
\begin{equation}
  \Gamma=\frac{2\pi}{\hbar}|M_{if}|^2\frac{dN}{dE_f}\,,
  \label{golden_rule}
\end{equation}
where $M_{if}\equiv\left\langle f|H_{\rm int}\right\rangle$
is the matrix element between initial and final states $i$ and $f$
with $H_{\rm int}$ the interaction energy, and $dN/dE_f$ is the final
state number density evaluated at the conserved total energy
of the final states.

As an example, we compute the rate $\Gamma$ for inverse $\beta$-decay
\begin{equation}
  \bar{\nu}_e+p\to n+e^+\,.\label{inverse_beta}
\end{equation}
We use the historical {\it Fermi theory} after which such
interactions are described by point-like couplings of four
fermions, symbolically $H_{\rm int}=G_{\rm F}\int d^3{\rm x}\psi^4$,
with Fermi's coupling constant $G_{\rm F}$. This yields
\begin{equation}
  \Gamma=\frac{2\pi}{\hbar}G_{\rm F}^2|M|^2\frac{dN}{dE_f}\,,
  \label{golden_rule2}
\end{equation}
where symbolically $M=\int d^3{\bf x}\psi^4$ which incorporates the
detailed structure of the interaction. If we normalize the volume $V$
to one, $M$ is dimensionless and of order unity, otherwise $M$ scales
as $V^{-1}$. In fact, it is
roughly the spin multiplicity factor, such that $|M|^2\simeq1$ if
the total leptonic angular momentum is 0, thus involving no
change of spin in the nuclei ("Fermi transitions"), whereas $|M|^2\simeq3$
if the total leptonic angular momentum is 1, thus involving
a change of spin in the nuclei ("Gamow-Teller transitions").
The final state density of a free particle is
\begin{equation}
  \frac{Vd^3{\bf p}}{(2\pi\hbar)^3}\,.
\end{equation}
Therefore, taking into account energy-momentum conservation, we get
in the center of mass (CM) frame the phase space factor for the two body
final state
\begin{equation}
  \frac{dN}{dE_f}=\int\frac{Vd^3{\bf p}_e}{(2\pi\hbar)^3}
  \frac{Vd^3{\bf p}_n}{(2\pi\hbar)^3}
  \frac{(2\pi)^3\delta^3({\bf p}_e+{\bf p}_n)}{V}
  \delta(E_e+E_n-E_0)\,,
\end{equation}
where ${\bf p}_e$, ${\bf p}_n$, $E_e$, $E_n$,
are momenta and kinetic energies of the electron and the
final state nucleus, respectively, and
$E_0$ is the total initial energy. Integrating out one of
the momenta gives $p_f\equiv p_e=p_n$ so that energy conservation
$E_0=(p_f^2+m_e^2)^{1/2}+(p_f^2+m_n^2)^{1/2}$ gives the factor
$dp_f/dE_0=(p_f/E_e+p_f/E_n)^{-1}=v_f^{-1}$ with $v_f$
being the relative velocity of the two final state particles.
This yields
\begin{equation}
  \frac{dN}{dE_f}=\frac{1}{2\pi^2}\frac{p_f^2}{v_f}\,.\label{phase_space3}
\end{equation}

We are now interested in the {\it cross section} $\sigma$ of the
two-body reaction Eq.~(\ref{inverse_beta}) defined by
\begin{equation}
  \Gamma=\sigma n_iv_i\,,\label{cross}
\end{equation}
where $n_i=V^{-1}$ and $v_i$ are density and velocity, respectively,
of one of the incoming particles in the frame where the other one
is at rest. Putting this together with Eqs.~(\ref{golden_rule2})
and (\ref{phase_space3}) finally yields
\begin{equation}
  \sigma(\bar{\nu}_ep\to ne^+)=\frac{G_{\rm F}^2}{\pi}|M_{if}|^2
  \frac{p_f^2}{v_iv_f}\,.\label{cross2}
\end{equation}
For $p_f\simeq1\,$MeV this cross section is $\sim10^{-43}\,{\rm cm}^2$.
In a target of proton density $n_p$ this gives a mean free path
defined by $l_\nu n_p\sigma(\bar{\nu}_ep\to ne^+)\sim1$. For water
this turns out to be $\sim30\,$pc which demonstrates the experimental
challenge associated with detection of neutrinos.

The first detections of this reaction was made by Reines and
Cowan in 1959. The source were neutron rich fission products
undergoing $\beta$-decay $n\to p e^-\bar{\nu}_e$. A 1000 MW reactor
gives a flux of $\sim10^{13}\,{\rm cm}^{-2}\,{\rm s}^{-1}$
$\bar{\nu}_e$s which they observed with a target of CdCl$_2$ and
water. Observed are fast electrons Compton scattered by annihilation
photons from the positrons within $\sim10^{-9}\,$s of the reaction
("prompt pulse") $\gamma-$rays from the neutrons captured by the
cadmium about $10^{-6}\,$s after the reaction ("delayed pulse").

\section{Dirac Fermions and the V--A Interaction}
Given the experimentally established fact that electroweak interactions
only involve left-handed neutrinos we now want to work out the
detailed structure of these interactions. In order to
do that we first have to introduce the Dirac fermion.

\subsection{Dirac Fermions as Representations of Space-Time Symmetries}
The {\it Poincar\'e group} is the symmetry group of special relativity
and consists of all transformations leaving invariant the metric
\begin{equation}
  ds^2=-(dx^0)^2+(dx^1)^2+(dx^2)^2+(dx^3)^2\,,\label{ds2}
\end{equation}
where $x^0$ is a time coordinate and $x^1$, $x^2$, and $x^3$ are
Cartesian space coordinates. These transformations are of the
form
\begin{equation}
  x^{\prime\mu}=\Lambda^\mu_\nu x^\nu+a^\mu\,,\label{poincare}
\end{equation}
where $a^\mu$ defines arbitrary space-time translations, and the
constant matrix $\Lambda^\mu_\nu$ satisfies
\begin{equation}
  \eta_{\mu\nu}\Lambda^\mu_\rho\Lambda^\nu_\sigma=\eta_{\rho\sigma}
  \,,\label{trafo_lambda}
\end{equation}
where $\eta_{\mu\nu}={\rm diag}(-1,1,1,1)$.
The unitary transformations on fields and physical states $\psi$ induced
by Eq.~(\ref{poincare}) satisfy the composition rule
\begin{equation}
  U(\Lambda_2,a_2)U(\Lambda_1,a_1)=U(\Lambda_2\Lambda_1,\Lambda_2a_1+a_2)
  \,.\label{composition}
\end{equation}
Important subgroups are defined by all elements with
$\Lambda=1$ (the commutative group of translations), and
by all elements with $a^\mu=0$ [the {\it homogeneous Lorentz
group} $SO(3,1)$ of matrices $\Lambda^\mu_\nu$ satisfying
Eq.~(\ref{trafo_lambda})]. The latter contains the subgroup
$SO(3)$ of all rotations for which $\Lambda^0_0=1$,
$\Lambda^\mu_0=\Lambda^0_\mu=0$ for $\mu=1,2,3$.

The general infinitesimal transformations of this type are characterized
by an anti-symmetric tensor $\omega^\mu_\nu$ and a vector $\epsilon^\mu$,
\begin{equation}
  \Lambda^\mu_\nu=\delta^\mu_\nu+\omega^\mu_\nu\,\quad
  a^\mu=\epsilon^\mu
  \,.\label{lambda}
\end{equation}
Any element $U(1+\omega,\epsilon)$ of the Poincar\'e group which is
infinitesimally close to the unit operator can then be expanded into
the corresponding hermitian generators $J^{\mu\nu}$ and $P^\mu$,
\begin{equation}
  U(1+\omega,\epsilon)=1+\frac{1}{2}i\omega_{\mu\nu}J^{\mu\nu}
  -i\epsilon_\mu P^\mu
  \,.\label{trafo1}
\end{equation}
It can be shown that these generators satisfy the commutation
relations
\begin{eqnarray}
  i\left[J^{\mu\nu},J^{\rho\sigma}\right]&=&\eta^{\nu\rho}J^{\mu\sigma}
  -\eta^{\mu\rho}J^{\nu\sigma}-\eta^{\sigma\mu}J^{\rho\nu}
  +\eta^{\sigma\nu}J^{\rho\mu}\nonumber\\
  i\left[P^\mu,J^{\rho\sigma}\right]
  &=&\eta^{\mu\rho}P^\sigma-\eta^{\mu\sigma}P^\rho
  \label{poincare_comm}\\
  \left[P^\mu,P^\nu\right]&=&0\,.\nonumber
\end{eqnarray}
The $P^\mu$ represent the energy-momentum vector, and since the
Hamiltonian $H\equiv P^0$ commutes with the spatial pseudo-three-vector
${\bf J}\equiv(J^{23},J^{31},J^{12})$, the latter represents the
angular-momentum which generates the group of rotations $SO(3)$.

The homogeneous Lorentz group implies that the dispersion relation
of free particles is of the form
\begin{equation}
  E^2(p)={\bf p}^2+M^2\,.\label{omegak}
\end{equation}
for a particle of mass $M$, momentum ${\bf p}$, and energy $E$.
If one now expands a free charged quantum field $\psi(x)$ into its
energy-momentum eigenfunctions
and interprets the coefficients $a({\bf p})$ of the positive energy solutions
as annihilator of a particle in mode ${\bf p}$, then the coefficients
$b^\dagger({\bf p})$ of the negative energy contributions have to be
interpreted as creators of anti-particles of opposite charge,
\begin{equation}
  \psi(x)=\sum_{{\bf p},E(p)>0}a({\bf p})u({\bf p})
  e^{-iE(k)t+i{\bf p}\cdot{\bf x}}
  +\sum_{{\bf p},E(p)<0}b^\dagger({\bf p})v({\bf p})
  e^{iE(p)t-i{\bf p}\cdot{\bf x}}\,.\label{expansion}
\end{equation}
Canonical quantization, shows that the creators and annihilators
indeed satisfy the relations,
\begin{equation}
  \left[a_i({\bf p}),a^\dagger_{i^\prime}({\bf p^\prime})\right]_\pm=
  \left[b_i({\bf p}),b^\dagger_{i^\prime}({\bf p^\prime})\right]_\pm=
  \delta_{ii^\prime}\delta({\bf p}-{\bf p^\prime})\,,\label{create_annihi}
\end{equation}
where $i,i^\prime$ now denote internal degrees of freedom such as spin,
and $[.,.]_\pm$ denotes the commutator for bosons, and the
anti-commutator for fermions, respectively.

Fields and physical states can thus be characterized by their
energy-momentum and spin, which characterize their transformation
properties under the group of translations and under the
rotation group, respectively. Let us first focus on fields and states
with non-vanishing mass. In this case one can perform a Lorentz boost
into the rest frame where $P^\mu=(M,0,0,0)$ with $M$ the mass of the
state. $P^\mu$ is then invariant under the rotation group $SO(3)$.
The irreducible unitary representations of this group are characterized
by a integer- or half-integer valued spin $j$ such that the $2j+1$ states are
characterized by the eigenvalues of $J_i$ which run over
$-j,-j+1,\cdots,j-1,j$. Note that an eigenstate with eigenvalue
$\sigma$ of $J_i$ is multiplied by a phase factor $e^{2\pi i\sigma}$
under a rotation around the $i-$axis by $2\pi$, and a half-integer
spin state thus changes sign. Given the fact that a rotation by
$2\pi$ is the identity this may at first seem surprising. Note,
however, that normalized states in quantum mechanics are only
defined up to phase factors and thus a general unitary {\it
projective representation} of a symmetry group on the Hilbert
space of states can in general include phase factors in the
composition rules such as Eq.~(\ref{composition}). This is indeed
the case for the rotation group $SO(3)$ which is isomorphic
to $S_3/Z_2$, the three-dimensional sphere in Euclidean four-dimensional
space with opposite points identified, and is thus {\it doubly
connected}. This means that closed curves winding $n$ times
over a closed path are continuously contractible to a point
if $n$ is even, but are not otherwise. Half-integer spins then
correspond to representations for which $U(\Lambda_1)U(\Lambda_2)=
(-)^nU(\Lambda_1\Lambda_2)$, where $n$ is the winding number
along the path from $1$ to $\Lambda_1$, to $\Lambda_1\Lambda_2$
and back to $1$, whereas integer spins do not produce a phase factor.

With respect to homogeneous Lorentz transformations, there are then
two groups of representations. The first one is formed by the tensor
representations which transform just as products of vectors,
\begin{equation}
  W^{\prime\mu\cdots}_{\nu\cdots}=\Lambda^\mu_\rho\Lambda_\nu^\sigma
  \cdots W^{\rho\cdots}_{\sigma\cdots}
  \,.\label{tensor}
\end{equation}
These represent bosonic degrees of freedom with maximal integer spin $j$
given by the number of indices. The simplest case is a complex
spin-zero scalar $\phi$ of mass $m$ whose standard free Lagrangian
\begin{equation}
  {\cal L}_\phi=-\frac{1}{2}
  \left(\partial_\mu\phi^\dagger\partial^\mu\phi-m^2\phi^\dagger\phi\right)
  \,,\label{L_phi}
\end{equation}
leads to an equation of motion known as {\it Klein-Gordon equation},
\begin{equation}
  \left(\partial_\mu\partial^\mu-m^2\right)\phi=0
  \,.\label{klein_gordon}
\end{equation}
In the static case $p^0=0$ this leads to an interaction potential
\begin{equation}
  V(r)=g_1g_2\frac{e^{-mr}}{r}\,,\label{potential}
\end{equation}
between two ``charges'' $g_1$ and $g_2$ which correspond to
sources on the right hand side of Eq.~(\ref{klein_gordon}). The
potential for the exchange of bosons of non-zero spin involve some
additional factors for the tensor structure. Note that the {\it range}
of the potential is given by $\simeq m^{-1}$. In the general case
$p^0\neq0$ the Fourier transform of Eq.~(\ref{klein_gordon}) with
a delta-function source term on the right hand side is
$\propto-i/(p^2+m^2)$. A four-fermion point-like interaction
of the form $G_{\rm F}\psi^4$ can thus be interpreted as the
low-energy limit $p^2\ll m^2$ of the exchange of a boson of
mass $m$. Later we will realize that the modern theory of
electroweak interactions is indeed based on the exchange of
heavy charged and neutral "gauge bosons". In the absence of
sources, Eq.~(\ref{klein_gordon}) gives the usual dispersion
relation $E^2=p^2+m^2$ for a free particle.

The second type of representation of the homogeneous Lorentz group
can be constructed from any set of {\it Dirac matrices} $\gamma^\mu$
satisfying the anti-commutation relations
\begin{equation}
  \left\{\gamma^\mu,\gamma^\nu\right\}=2\eta^{\mu\nu}
  \,,\label{clifford}
\end{equation}
also known as {\it Clifford algebra}. One can then show that
the matrices
\begin{equation}
  J^{\mu\nu}\equiv-\frac{i}{4}\left[\gamma^\mu,\gamma^\nu\right]
  \label{jmunu_spinor}
\end{equation}
indeed obey the commutation relations in Eq.~(\ref{poincare_comm}).
The objects on which these matrices act are called {\it Dirac spinors}
and have spin $1/2$. In 3+1 dimensions, the smallest representation
has four complex components, and thus the $\gamma^\mu$ are $4\times4$
matrices. A possible representation of Eq.~(\ref{clifford}) is
\begin{equation}
  \gamma_i=\left(\matrix{0& -i\sigma_i\cr i\sigma_i& 0}\right)
  \,,\quad i=1,2,3\,,
  \quad\gamma_0=i\left(\matrix{1& 0\cr 0& -1}\right)\,,\label{gammamunu}
\end{equation}
where $\sigma_i$ are the Pauli matrices.

The standard free Lagrangian for a spin-$1/2$ Dirac spinor
$\psi$ of mass $m$,
\begin{equation}
  {\cal L}_\psi=-\bar\psi(\gamma^\mu\partial_\mu+m)\psi\,,\label{L_psi}
\end{equation}
where $\bar\psi\equiv\psi^\dagger i\gamma^0$,
leads to an equation of motion known as {\it Dirac equation},
\begin{equation}
  \left(\gamma^\mu\partial_\mu+m\right)\psi=0
  \,.\label{dirac}
\end{equation}
Its free solutions also satisfy the Klein-Gordon equation
Eq.~(\ref{klein_gordon}) and are of the form Eq.~(\ref{expansion})
where, up to a normalization factor $N$,
\begin{equation}
  u({\bf p})=N
  \left({\matrix{\tilde{u}\cr\frac{\ss\cdot{\bf p}}{E+m}\tilde{u}}}\right)
  \,,\quad
  v({\bf p})=N
  \left({\matrix{\frac{\ss\cdot{\bf p}}{E+m}\tilde{v}\cr\tilde{v}}}\right)
  \,.
\end{equation}
Here, $u$ and $v$ are 4-spinors, whereas $\tilde{u}$ and $\tilde{v}$
are two-spinors.

It is easy to see that the matrix
\begin{equation}
  \gamma_5\equiv-i\gamma^0\gamma^1\gamma^2\gamma^3=
  -\left(\matrix{0&1\cr1&0}\right)\label{gamma_5}
\end{equation}
is a pseudo-scalar because the spatial $\gamma^i$ change sign under
parity transformation, and satisfies
\begin{equation}
  \gamma_5^2=1\,\quad\{\gamma^\mu,\gamma_5\}=0\,
  \quad [J^{\mu\nu},\gamma_5]=0\,.\label{gamma_5_prop}
\end{equation}
A four-component Dirac spinor $\psi$ can then be split into two inequivalent
{\it Weyl representations} $\psi_L$ and $\psi_R$ which are called
left-chiral and right-chiral,
\begin{equation}
  \psi=\psi_L+\psi_R\equiv\frac{1+\gamma_5}{2}\psi+\frac{1-\gamma_5}{2}\psi
  \,.\label{chirality}
\end{equation}
Note that according to Eqs.~(\ref{gamma_5_prop}) and~(\ref{chirality})
the mass term in the Lagrangian Eq.~(\ref{L_psi}) flips chirality,
whereas the kinetic term conserves chirality.

The general irreducible representations of the
homogeneous Lorentz group are then given by arbitrary direct
products of spinors and tensors. We note that massless states
form representations of the group $SO(2)$ leaving invariant $P^\mu$,
instead of of $SO(3)$. The group $SO(2)$ has only one generator
which can be identified with {\it helicity}, the projection of spin onto
three-momentum. For fermions this is the chirality defined by $\gamma_5$
above.

In the presence of mass the relation between chirality and helicity
$H\equiv\ss\cdot{\bf p}/p$ is more complicated:
\begin{eqnarray}
  \frac{1\pm\gamma_5}{2}u({\bf p})&=&\frac{N}{2}
  \left(1\mp\frac{\ss\cdot{\bf p}}{E+m}\right)
  \left(\matrix{\tilde{u}\cr\mp\tilde{u}}\right)\label{chirhel}\\
  &=&\frac{N}{2}
  \left[\left(1\mp\frac{p}{E+m}\right)
  \frac{1+H}{2}+\left(1\pm\frac{p}{E+m}\right)\frac{1-H}{2}\right]
  \left(\matrix{\tilde{u}\cr\mp\tilde{u}}\right)
  \nonumber\\
  \frac{1\pm\gamma_5}{2}v({\bf p})&=&
  \mp\frac{N}{2}
  \left[\left(1\mp\frac{p}{E+m}\right)
  \frac{1+H}{2}+\left(1\pm\frac{p}{E+m}\right)\frac{1-H}{2}\right]
  \left(\matrix{\tilde{v}\cr\mp\tilde{v}}\right)
  \,.\nonumber
\end{eqnarray}
From this follows that in a chiral state $u_{L,R}$ the helicity
polarization is given by
\begin{equation}
  P_{L,R}=\frac{I_+^{L,R}-I_-^{L,R}}{I_+^{L,R}+I_-^{L,R}}=
  \mp\frac{p}{E}\,,\label{chirhel2}
\end{equation}
where $I_\pm^{L,R}$ are the intensities in the $H=\pm1$ states for
given chirality $L$ or $R$. Note that
due to Eq.~(\ref{expansion}) the physical momentum of
anti-particles described by the $v$ spinor is $-{\bf p}$ in this
convention, and therefore the helicity polarization for anti-particles
in pure chiral states are opposite from Eq.~(\ref{chirhel2}):
Left chiral particles are predominantly left-handed and left-chiral
anti-particles are predominantly right-handed in the relativistic
limit. Furthermore, helicity and chirality commute exactly only in the
limit $p\gg m$, $v\to1$. The experimental fact that observed electron
and neutrino helicities are $\mp v$ for particles and anti-particles,
respectively, where $v$ is the particle velocity, now implies
that both electrons and neutrinos and their anti-particles are fully
left-chiral.

\subsection{The $V$--$A$ Coupling}
Since Dirac spinors have 4 independent components, there are
16 independent bilinears listed in Tab.~\ref{tab2}. Using the
equality
\begin{equation}
  \gamma_\mu^\dagger=\gamma_0\gamma_\mu\gamma_0\,,\label{dagger}
\end{equation}
which can easily be derived from Eq.~(\ref{gammamunu}), one
sees that the phase factors of the bilinears in Tab.~\ref{tab2}
are chosen such that their hermitian conjugate is the same
with $\psi_1\leftrightarrow\psi_2$.

\begin{table}[ht]
\caption{The Dirac bilinears. For $\psi_1=\psi_2$ these are real.}
\label{tab2}
\begin{tabular}{ccc}
$\bar\psi_1\psi_2$ & scalar & $S$ \\
$i\bar\psi_1\gamma_\mu\psi_2$ & 4-vector & $V$ \\
$i\bar\psi_1\gamma_\mu\gamma_\nu\psi_2$ & tensor & $T$ \\
$i\bar\psi_1\gamma_\mu\gamma_5\psi_2$ & axial 4-vector & $A$ \\
$i\bar\psi_1\gamma_5\psi_2$ & pseudo-scalar & $P$ \\
\end{tabular}
\end{table}

Lorentz invariance implies that the matrix element of a general
$\beta$-interaction is of the form
\begin{equation}
  M=G_{\rm F}\sum_{i=S,V,T,A,P}C_i(\bar\psi_1{\cal O}_i\psi_2)
  (\bar\psi_3{\cal O}_i\psi_4)\,,\label{mew}
\end{equation}
such that only the same types of operators ${\cal O}_i$ from
Tab.~\ref{tab2} couple and common Lorentz indices are contracted
over.

Eq.~(\ref{mew}) is a Lorentz scalar. However, we know that electroweak
interactions violate parity and thus we have to add pseudo-scalar
quantities to Eq.~(\ref{mew}). Equivalently, we can substitute
any lepton spinor $\psi$ in Eq.~(\ref{mew}) by
$\frac{1}{2}(1+\gamma_5)\psi$. This is correct at least for the
interactions with charge exchange, the so called {\it charged current}
interactions, for which we know experimentally that both neutrinos and
charged leptons are fully left-chiral. Using Eq.~(\ref{gamma_5_prop}),
this leads to terms of the form
\begin{equation}
  \overline{l_{i,L}^-}{\cal O}\nu_{i,L}=
  \overline{l_i^-}(1-\gamma_5){\cal O}(1+\gamma_5)\nu_i\,,\quad l=e,\mu,\tau\,,
  \label{dirac_bilin}
\end{equation}
which implies that only the $V$ and $A$ type interactions from
Tab.~\ref{tab2} can contribute. The general form of charged
current interactions involving neutrinos is therefore usually
written as
\begin{equation}
  M_{\rm cc}^\nu=\frac{G_{\rm F}}{\sqrt2}
  \left[\bar\psi_1\gamma^\mu(C_V+C_A\gamma_5)\psi_2\right]
  \left[\overline{l_i^-}\gamma_\mu(1+\gamma_5)\nu_i\right]\,,\label{mcc}
\end{equation}

\section{Divergences in the Weak Interactions and Renormalizability}
\label{sec_renorm}
A incoming plane wave $\psi_i\equiv e^{ikz}$ of momentum $k$ in the
$z$-direction can be
expanded into incoming and outgoing radial modes $e^{-ikr}$ and
$e^{ikr}$, respectively, in the following way
\begin{equation}
  e^{ikz}=\frac{i}{2kr}\sum_l(2l+1)\left[(-1)^l e^{-ikr}-e^{ikr}\right]
  P_l(\cos\theta)\,,\label{planewave}
\end{equation}
where $P_l(x)$ are the Legendre polynomials and $\cos\theta=z/r$.
Scattering modifies the outgoing modes by multiplying them with a
phase $e^{2i\delta_l}$ and an amplitude $\eta_l$ with $0\leq\eta_l\leq1$.
The scattered outgoing wave thus has the form
\begin{equation}
  \psi_{\rm scatt}=\frac{e^{ikr}}{kr}\sum_l(2l+1)
  \frac{\eta_l e^{2i\delta_l}-1}{2i}\,P_l(\cos\theta)
  \equiv\frac{e^{ikr}}{r}\,F(\theta)\,,\label{psiscatt}
\end{equation}
where $F(\theta)$ is called the scattering amplitude.

Let us now imagine elastic scattering in the CM frame, where momentum
$p_*$ and velocity $v$ are equal before and after scattering. The incoming
flux is then $v|\psi_i|^2=v$ and the outgoing flux through a solid angle
$d\Omega$ is $v|\psi_{\rm scatt}|^2 r^2d\Omega=v|F(\theta)|^2d\Omega$.
The definition Eq.~(\ref{cross}) of the scattering cross section then yields
\begin{equation}
  \left(\frac{d\sigma}{d\Omega}\right)_{\rm el}=|F(\theta)|^2\,.\label{elcross}
\end{equation}
Using orthogonality of the Legendre polynomials,
$\int d\Omega P_l(\Omega)P_{l^\prime}(\Omega)=4\pi\delta_{ll^\prime}/(2l+1)$,
in Eq.~(\ref{psiscatt}), we obtain for the total elastic scattering
cross section
\begin{equation}
  \sigma_{\rm el}=\frac{4\pi}{p_*^2}\sum_l(2l+1)
  \left|\frac{\eta_le^{2i\delta_l}-1}{2i}\right|^2\,.\label{elcrosstot}
\end{equation}
For scattering of waves of angular momentum $l$ this results in the upper
limit
\begin{equation}
  \sigma_{{\rm el},l}\leq\frac{4\pi}{p_*^2}(2l+1)
  \,,\label{unitarity}
\end{equation}
which is called {\it partial wave unitarity}.

On the other hand, in Fermi theory typical cross sections grow with $p_*$
as in Eq.~(\ref{cross2}) and violate Eq.~(\ref{unitarity}) for s-waves
($l=0$) for
\begin{equation}
  \frac{4G_{\rm F}^2p_*^2}{\pi}\ga\frac{4\pi}{p_*^2}\,,\label{swave}
\end{equation}
where we used $\sum|M_{if}|^2\simeq4$ for the sum over polarizations.
This occurs for $p_*\ga(\pi/G_{\rm F})^{1/2}\simeq500\,$GeV. Such energies
are nowadays routinely achieved at accelerators such as in the Tevatron
at Fermilab. As will be seen in the next section, this is ultimately
due to the fact that the coupling constant $G_{\rm F}$ has negative
energy dimension and corresponds to a {\it non-renormalizable} interaction.
This will be cured by spreading the contact interaction with the propagator
of a gauge boson of mass $M\simeq G_{\rm F}^{-1/2}\sim300\,$GeV. This
corresponds to multiplying the l.h.s. of Eq.~(\ref{swave}) with
the square of the propagator, $\simeq(1+p_*^2/M^2)^{-2}$, which thus
becomes $4/(\pi p_*^2)$ for $p_*\to\infty$. This scaling with $p_*$
is of course a simple consequence of dimensional analysis.
As a result, partial wave
unitarity is not violated any more at high energies, at least within
this rough order of magnitude argument. The gauge theory of electroweak
interactions discussed below is {\it renormalizable}.

Theories which contain only coupling terms of non-negative mass
dimension lead to
only a finite number of graphs diverging at large energies. It turns
out that these divergences can be absorbed into the finite number of parameters
of the theory which is why they are called {\it renormalizable}.

Good examples of non-renormalizable interaction terms are
given by
\begin{equation}
  \frac{ie}{2M}\bar{\psi}[\gamma_\mu,\gamma_\nu]\psi F^{\mu\nu}\,,
  \quad
  \frac{e}{2M}\bar{\psi}\gamma_5[\gamma_\mu,\gamma_\nu]\psi F^{\mu\nu}
  \,,\label{moments}
\end{equation}
where $M$ is some large mass scale presumably related to grand
unification and $e$ is the (positive) electric
charge unit. The gauge invariant field strength tensor
$F_{\mu\nu}=\partial_\mu A_\nu-\partial_nu A_\mu$ in terms of
the gauge potential Eq.~(\ref{var_A}) below represents the electric
field strength $E^i=-\partial_0 A^i-\partial_i A^0=F^{0i}$ and
magnetic field strength $B^i=\epsilon^{ijk}\partial_j A_k=\epsilon^{ijk}F_{jk}$
where Latin indices represent spatial indices and $\epsilon^{ijk}$ is
totally anti-symmetric with $\epsilon^{123}=1$. As a consequence,
in the non-relativistic limit, Eq.~(\ref{moments}) reduce to a
magnetic and electric dipole moment of the $\psi$ field, respectively,
of size $4e/M$. Note that these are even and odd, respectively,
under parity and time reversal.

Other non-renormalizable terms may arise from Lorentz symmetry violation
by physics close to the grand unification scale $M$. In Sect.~\ref{sec_vli}
we will see how high energy astrophysics can constrain such terms and
thus physics beyond the Standard Model to precisions greater than
laboratory experiments.

\section{Gauge Symmetries and Interactions}
\subsection{Symmetries of the Action}
Lorentz invariance suggests that the action should be the space-time
integral of a scalar function of the fields $\psi_i({\bf x},t)$ and
their space-time derivatives $\partial_\mu\psi_i({\bf x},t)$,
and thus that the Lagrangian should be the space-integral of a scalar
called the {\it Lagrangian density} ${\cal L}$,
\begin{equation}
  S[\psi]=\int d^4x{\cal L}[\psi_i(x),\partial_\mu\psi_i(x)]\,,
  \label{lagrangian}
\end{equation}
where $x\equiv({\bf x},t)$ from now on. In this case, the equations
of motion read
\begin{equation}
  \partial_\mu\frac{\partial{\cal L}}{\partial(\partial_\mu\psi_i)}=
  \frac{\partial{\cal L}}{\partial\psi_i}\,,\label{eq_motion2}
\end{equation}
which are called {\it Euler-Lagrange equations} and are obviously
Lorentz invariant if ${\cal L}$ is a scalar.

Symmetries can be treated in a very transparent way in the Lagrangian
formalism. Assume that the action is invariant, $\delta S=0$, independent
of whether $\psi_i(x)$ satisfy the field equations or not,
under a global symmetry transformation,
\begin{equation}
  \delta\psi_i(x)=i\epsilon{\cal F}_i[\psi_j(x),\partial_\mu\psi_j(x)]
  \,,\label{global_sym}
\end{equation}
for which $\epsilon$ is independent of $x$. Here and in the following
explicit factors of $i$ denote the imaginary unit, and not an index.
Then, for a space-time dependent $\epsilon(x)$, the variation must be
of the form
\begin{equation}
  \delta S=-\int d^4x J^\mu[x,\psi_j(x),\partial_\mu\psi_j(x)]
  \partial_\mu\epsilon(x)\,.\label{delta_S}
\end{equation}
But if the fields satisfy their equations of motion, $\delta S=0$,
and thus
\begin{equation}
  \partial_\mu J^\mu[x,\psi_j(x),\partial_\mu\psi_j(x)]=0
  \,,\label{noether}
\end{equation}
which implies Noethers theorem, the existence of one conserved
current $J^\mu$ for each continuous global symmetry.
If Eq.~(\ref{global_sym}) leaves the Lagrangian density itself
invariant, an explicit formula for $J^\mu$ follows immediately,
\begin{equation}
  J^\mu=-i\frac{\partial{\cal L}}{\partial(\partial_\mu\psi_i)}
  {\cal F}_i\,,\label{j_explicit}
\end{equation}
where we drop the field arguments from now on.

As opposed to a global symmetry, Eq.~(\ref{global_sym}), which
leaves a theory invariant under a transformation that is the
same at all space-time points, a gauge symmetry is more powerful
as it leaves invariant a theory, i.e. $\delta{\cal L}=\delta S=0$,
under transformations that can
be chosen independently at each space-time point. Gauge symmetries
are usually also linear in the (fermionic) matter fields which we represent
here by one big spinor $\psi(x)$ that in general contains Lorentz
spinor indices as well as some internal group indices on which the
gauge transformations act. For real infinitesimal $\epsilon^\alpha(x)$
we write
\begin{equation}
  \delta\psi(x)=i\epsilon^\alpha(x)t_\alpha\psi(x)
  \,,\label{gauge_sym1}
\end{equation}
where $\alpha$ labels the different independent generators $t_\alpha$
of the gauge group. A finite gauge transformation would be written
as $\psi(x)\to\exp(i\Lambda^\alpha(x)t_\alpha)\psi(x)$ and reduces to
Eq.~(\ref{gauge_sym1}) in the limit
$\Lambda^\alpha(x)=\epsilon^\alpha(x)\to0$. The hermitian matrices
$t_\alpha$ form a {\it Lie algebra} with commutation relations
\begin{equation}
  [t_\alpha,t_\beta]=iC^\gamma_{\alpha\beta}t_\gamma
  \,,\label{commutation}
\end{equation}
where the real constants $C^\gamma_{\alpha\beta}$ are called
structure constants of the Lie algebra, and are anti-symmetric in
$\alpha\beta$.

\subsection{Gauge Symmetry of Matter Fields}
If the Lagrangian contained no field derivatives, but only terms
of the form $\psi(x)^\dagger\cdots\psi(x)$, there would be no
difference between global and local gauge invariance. However,
dynamical theories contain space-time derivatives $\partial_\mu\psi$
which transform differently under Eq.~(\ref{gauge_sym1}) than
$\psi$, and thus would spoil local gauge invariance. One can
cure this by introducing new vector {\it gauge fields} $A^\alpha_\mu(x)$
and defining {\it covariant derivatives} by
\begin{equation}
  D_\mu\psi(x)\equiv\partial_\mu\psi(x)-iA^\alpha_\mu(x)t_\alpha\psi(x)
  \,.\label{covariant}
\end{equation}
The gauge variation of this from the variation of $\psi$ alone (i.e. assuming
$A^\alpha_\mu$ constant for the moment) reads
\begin{equation}
  \delta_\psi D_\mu\psi(x)=i\epsilon^\alpha(x)t_\alpha D_\mu\psi(x)
  +i\left[\partial_\mu\epsilon^\alpha(x)+
  C^\alpha_{\beta\gamma}A^\beta_\mu(x)\epsilon^\gamma(x)\right]
  t_\alpha\psi(x)\,,\label{var_covariant}
\end{equation}
where we have used Eq.~(\ref{commutation}). The new term proportional to
the structure constants $C^\alpha_{\beta\gamma}$ results from moving
the gauge variation of $\psi$ in Eq.~(\ref{covariant}) to the
left of the matrix gauge field $A^\alpha_\mu(x)t_\alpha$ and is
only present in {\it non-abelian gauge theories} for which the
$t_\alpha$ do not all commute. The variation
$\delta_\psi S_{\rm m}$ of the matter action $S_{\rm m}$ can then
be obtained from Eq.~(\ref{delta_S}), generalized to several
$\epsilon^\alpha$, and with $\partial_\mu\epsilon^\alpha(x)$ substituted
by the corresponding first factor of the second term in
Eq.~(\ref{var_covariant}),
\begin{equation}
  \delta_\psi S_{\rm m}=-\int d^4x J^\mu_\alpha(x)
  \left[\partial_\mu\epsilon^\alpha(x)+
  C^\alpha_{\beta\gamma}A^\beta_\mu(x)\epsilon^\gamma(x)\right]
  \,,\label{var_Sm}
\end{equation}
where the {\it gauge currents}
Eq.~(\ref{j_explicit}) now read [compare Eqs.~(\ref{global_sym})
and~(\ref{gauge_sym1})]
\begin{equation}
  J^\mu_\alpha=-i\frac{\partial{\cal L}_{\rm m}}
  {\partial(\partial_\mu\psi)}t_\alpha\psi\label{gauge_current}
\end{equation}
in terms of the matter Lagrangian ${\cal L}_{\rm m}$. Realizing now that
$[\partial S_{\rm m}/\partial A^\alpha_\mu(x)]=[\partial{\cal L}_{\rm m}/
\partial(\partial_\mu\psi)](-it_\alpha\psi)=J^\mu_\alpha$, we
see that $\delta S_{\rm m}=\delta_\psi S_{\rm m}+
[\partial S_{\rm m}/\partial A^\alpha_\mu(x)]\delta A^\alpha_\mu(x)$
vanishes identically if we adopt the gauge transformation
\begin{equation}
  \delta A^\alpha_\mu(x)=\partial_\mu\epsilon^\alpha(x)+
  C^\alpha_{\beta\gamma}A^\beta_\mu(x)\epsilon^\gamma(x)
  \,,\label{var_A}
\end{equation} 
for the gauge field $A^\alpha_\mu(x)$.

The standard gauge-invariant term for fermions is then given by
the matter Lagrange density
\begin{equation}
  {\cal L}_{\rm m}=-\bar\psi(\gamma^\mu D_\mu+m)\psi
  =-\bar\psi(\gamma^\mu\partial_\mu+m)\psi+A^\alpha_\mu J^\mu_\alpha
  \,,\label{L_matter}
\end{equation}
where $m$ is the fermion mass matrix. The second equality shows
how the matter Lagrangian splits into the free part quadratic
in the fields, Eq.~(\ref{L_psi}), and the fundamental coupling
of the gauge field to the gauge current Eq.~(\ref{gauge_current}).
Since ${\cal L}_{\rm m}$ is real and the gauge current is hermitian,
$J_{\mu\alpha}=J^\dagger_{\mu\alpha}$, the gauge fields $A^\alpha_\mu$
are also real.

\subsection{Gauge Theory of the Electroweak Interaction}
In the electroweak Standard Model the elementary fermions are arranged
into three families or generations which here are labeled with the index $i$.
Each family consists of a left-chiral doublet of leptons,
$\left(\matrix{\nu_i\cr l^-_i}\right)_L$, a left-chiral doublet of quarks,
$\left(\matrix{u_i\cr d_i}\right)_L$, and the corresponding
right-chiral singlets $l^-_{iR}$, $u_{iR}$, and $d_{iR}$. Here,
left- and right-chiral is understood
as in Eq.~(\ref{chirality}), and each quark species comes in three
colors corresponding to the three-dimensional representations
of the strong interaction gauge group $SU(3)$ whose index is suppressed
here. The three known leptons are the electron, muon, and tau
with their corresponding neutrinos. The three up-type quarks
are called up, charm-, and top-quark, and the down-type quarks
are the down-, strange-, and bottom-quarks. The fermion masses
rise steeply with generation from about 1 MeV for the first generation
to up to $\simeq175\,$GeV for the top-quark whose direct discovery
occurred as late as 1995 at Fermilab in the USA.

Note that no right-handed neutrino appears and thus neutrino mass terms of
the form $\overline{\nu_L}\nu_R+$h.c. (h.c. denotes hermitian conjugate
here and in the following) are absent in the Standard Model.
Implications of recent experimental evidence for neutrino masses
for modifications of the Standard Model will not be discussed here.
To simplify the notation we assemble all fields into lepton and quark
doublets, $l_i\equiv\left(\matrix{\nu_i\cr l^-_i}\right)$, and
$q_i\equiv\left(\matrix{u_i\cr d_i}\right)$, including the right-handed
components. We will also use the
Pauli matrices
\begin{eqnarray}
  (\tau_0,\mbox{\boldmath$\tau$})&=&(\tau_0,\tau_1,\tau_2,\tau_3)
  \label{pauli}\\
  &\equiv&\left\{\left(\matrix{1 & 0\cr 0 & 1}\right),
  \left(\matrix{0 & 1/2\cr 1/2 & 0}\right)\,,
  \left(\matrix{0 & -i/2\cr i/2 & 0}\right)\,,
  \left(\matrix{1/2 & 0\cr 0 & -1/2}\right)\right\}\,.\nonumber
\end{eqnarray}

The electroweak gauge group is given by
\begin{equation}
  G=SU(2)_L\times U(1)_Y\,,\label{ew}
\end{equation}
where the first factor only acts on the left-handed doublets. Denoting
the dimensionless coupling constants corresponding to these two factors
with $g$ and $g^\prime$, we write the four generators in the leptonic
and quark sector as
\begin{eqnarray}
  {\bf t}_l={\bf t}_q&\equiv&(t_1,t_2,t_3)
  \equiv g\frac{1+\gamma_5}{2}\,\mbox{\boldmath$\tau$}\nonumber\\
  t_{Yl}&=&g^\prime\left[\frac{1+\gamma_5}{2}\frac{\tau_0}{2}+
  \frac{1-\gamma_5}{2}\tau_0\right]\label{t_ew}\\
  t_{Yq}&=&g^\prime\left[-\frac{1+\gamma_5}{2}\frac{\tau_0}{6}
  -\frac{1-\gamma_5}{2}\left(\frac{\tau_0}{6}+\tau_3\right)\right]
  \,.\nonumber
\end{eqnarray}
These correspond to the generators $t_\alpha$ from the previous
section, and we denote the corresponding gauge fields by
${\bf A}_\mu$ and $B_\mu$. It is easy to see that the electric charge
operator is then given by the combination
\begin{equation}
  q=\frac{e}{g}\,t_3-\frac{e}{g^\prime}\,t_Y\,,\label{echarge}
\end{equation}
where $e$ is the (positive) electric charge unit.

We are here only interested in the part of the Lagrangian involving
matter fields. This is then given by Eq.~(\ref{L_matter}) where
$\psi$ now represents all lepton and quark multiplets $l_i$ and $q_i$.
Using Eq.~(\ref{covariant}), where, from comparing Eq.~(\ref{t_ew})
with Eq.~(\ref{commutation}),
$C^\alpha_{\beta\gamma}=g\epsilon_{\alpha\beta\gamma}$ for $SU(2)_L$,
and zero for $U(1)$, we can write the matter part of the electroweak
Lagrangian as
\begin{eqnarray}
  {\cal L}_{\rm ew,m}&=&-\sum_{i=1}^3\bar{l}_i\gamma^\mu\left(\partial_\mu
  -i{\bf A}_\mu\cdot{\bf t}-iB_\mu t_{Yl}\right)l_i\label{L_ew}\\
  &&-\sum_{i=1}^3\bar{q}_i\gamma^\mu\left(\partial_\mu
  -i{\bf A}_\mu\cdot{\bf t}-iB_\mu t_{Yq}\right)q_i
  \,.\nonumber
\end{eqnarray}

It will be more convenient to use charge eigenstates as basis of
the electroweak gauge bosons and to identify the photon $A_\mu$ as
carrier of the electromagnetic interactions. There is then one other
neutral gauge boson $Z_\mu$ and two gauge bosons $W^\pm$ of charge
$\pm e$. They are defined by
\begin{eqnarray}
  A^1_\mu&=&\frac{1}{\sqrt2}\left(W^-_\mu+W^+_\mu\right)\nonumber\\
  A^2_\mu&=&\frac{1}{\sqrt2}\left(W_\mu^--W^+_\mu\right)\label{wwza}\\
  A^3_\mu&=&\cos\theta_{\rm ew}Z_\mu-\sin\theta_{\rm ew}A_\mu\nonumber\\
  B_\mu&=&\sin\theta_{\rm ew}Z_\mu+\cos\theta_{\rm ew}A_\mu\,,\nonumber\\
\end{eqnarray}
where the {\it electroweak angle} $\theta_{\rm ew}$ is defined by
\begin{equation}
  g=\frac{e}{\sin\theta_{\rm ew}}\,,\quad
  g^\prime=\frac{e}{\cos\theta_{\rm ew}}\,.\label{theta_ew}
\end{equation}
The interaction terms in Eq.~(\ref{L_ew}) can then be written as
\begin{eqnarray}
  {\bf A}_\mu\cdot{\bf t}+B_\mu t_Y&=&\frac{g}{\sqrt2}\frac{1+\gamma_5}{2}
  \left(W^+_\mu\tau^++W^-_\mu\tau^-\right)\label{wwza2}\\
  &&-\frac{g}{2\cos\theta_{\rm ew}}
  Z_\mu\left(\frac{1+\gamma_5}{2}\tau_3
  -q\sin^2\theta_{\rm ew}\right)+A_\mu q\,,\nonumber
\end{eqnarray}
where $\tau^\pm\equiv\tau_1\pm i\tau_2$ are the weak isospin
raising and lowering operators, respectively.

Up to this point all fields are massless. Mass terms for gauge bosons
and for fermions, whether Dirac or Majorana (see below), are inconsistent
with gauge invariance. The standard way to introduce them is by
{\it spontaneously broken gauge symmetries}. Without going into
any detail here, we just mention that this is done by introducing
a scalar {\it Higgs} field coupling to gauge boson and fermion
bilinears in a gauge-invariant way and making it adopt a vacuum
expectation value due to a suitably chosen potential.

Let us now consider processes involving the exchange of a
$W^\pm$ or $Z$ boson with energy-momentum transfer $q$ much
smaller than the gauge boson mass, $|q^2|\ll m^2_{W,Z}$, such that
the boson propagator can be approximated by $-i\eta_{\mu\nu}/m^2_{W,Z}$.
In this case the second order terms in the perturbation series
give rise to {\it effective interactions} of the form
\begin{equation}
  \frac{1}{m^2_W}J^\mu_{\rm cc}J^\dagger_{\mu{\rm cc}}+
  \frac{1}{m^2_Z}J^\mu_{\rm nc}J_{\mu{\rm nc}}
  \,.\label{ccnc}
\end{equation}
Here, the {\it charged current} and {\it neutral current}
are gauge currents given by comparing Eq.~(\ref{L_matter}) with
Eq.~(\ref{L_ew}), and using Eq.~(\ref{wwza2}),
\begin{eqnarray}
  J^\mu_{\rm cc}&=&i\frac{g}{\sqrt2}\sum_{i=1}^3\left[
  \bar{l}_i\gamma^\mu\frac{1+\gamma_5}{2}\tau^+l_i
  +\bar{q}_i\gamma^\mu\frac{1+\gamma_5}{2}\tau^+q_i\right]\label{ccnc2}\\
  J^\mu_{\rm nc}&=&-i\frac{g}{2\cos\theta_{\rm ew}}\sum_{i=1}^3
  \biggl[\bar{l}_i\gamma^\mu\left(
  \frac{1+\gamma_5}{2}\tau_3-q\sin^2\theta_{\rm ew}\right)l_i\nonumber\\
  &&\hspace{2.6cm}+\bar{q}_i\gamma^\mu\left(\frac{1+\gamma_5}{2}\tau_3
  -q\sin^2\theta_{\rm ew}\right)q_i
  \biggr]\,.\nonumber
\end{eqnarray}
Eqs.~(\ref{ccnc}) and~(\ref{ccnc2}) provide an effective description
of all low energy weak processes. This is an instructive example
of how a more fundamental renormalizable description of interactions
at high energies, in this case electroweak gauge theory, can reduce to
an effective non-renormalizable description of interactions at low
energies which are suppressed by a large mass scale, in this case
$m_W$ or $m_Z$. In fact, the latter is identical in form with the
historical ``V--A'' theory Eq.~(\ref{mcc}) which, for example, for
the muon decay $\mu^-\to e^-\bar\nu_e\nu_\mu$ reads
\begin{equation}
  \frac{G_{\rm F}}{\sqrt2}\left[\overline{e^-}\gamma^\mu(1+\gamma_5)
  \nu_e\right]
  \left[\bar{\nu}_\mu\gamma_\mu(1+\gamma_5)\mu^-\right] +\mbox{h.c.}
  \,,\label{v_a}
\end{equation}
where the {\it Fermi constant} $G_{\rm F}$ by comparison with
Eqs.~(\ref{ccnc}) and~(\ref{ccnc2}) is given by
\begin{equation}
  G_{\rm F}=\frac{g^2}{4\sqrt2 m^2_W}=
  1.16637(1)\times10^{-5}\,{\rm GeV}^{-2}\,.\label{gf}
\end{equation}

Radioactive $\beta-$decay processes are described by the terms in
Eq.~(\ref{ccnc}) containing $\overline{e^-}\gamma_\mu(1+\gamma_5)\nu_e$
or its hermitian conjugate for one of the charged currents
$J_{\mu\rm cc}$ or $J^\dagger_{\mu\rm cc}$, and a quark term
for the other current. For example, neutron decay, $n\to pe^-\bar\nu_e$
is due to the contribution $\bar{u}\gamma^\mu(1+\gamma_5)d$ to
$J^\mu_{\rm cc}$, which causes one of the d-quarks in the neutron
to transform into a u-quark under emission of a $W^-$ boson
which in turn decays into $e^-\bar\nu_e$, $(udd)\to(uud)e^-\bar\nu_e$.

Inverting Eq.~(\ref{wwza}) to $Z_\mu=\cos\theta_{\rm ew}A_\mu^3+
\sin\theta_{\rm ew}B_\mu$ and writing out the mass term of the
neutral gauge boson sector $\frac{1}{2}m_Z^2Z_\mu Z^\mu$ implies
\begin{equation}
m_W=m_Z\cos\theta_{\rm ew}\label{ewmass}\,,
\end{equation}
because the mass term of $A_\mu^3$ has to be identical to the
one for $W_\mu^\pm$. From this it follows immediately that the
neutral current part of Eqs.~(\ref{ccnc}), (\ref{ccnc2}) involving
neutrinos can be written as
\begin{equation}
  \frac{G_{\rm F}}{\sqrt2}\left[\bar{\nu}_i\gamma^\mu(1+\gamma_5)
  \nu_i\right]
  \left[\bar{\psi}\gamma_\mu\left(g_L(1+\gamma_5)+g_R(1-\gamma_5)\right)
  \psi\right]\,,\label{v_a2}
\end{equation}
where $\psi$ stands for quarks and leptons and
\begin{eqnarray}
  g_L&=&\tau_3-q\sin^2\theta_{\rm ew}\nonumber\\
  g_R&=&-q\sin^2\theta_{\rm ew}\,.\label{glr}
\end{eqnarray}

\section{Neutrino Scattering}

\begin{figure}
\includegraphics[width=0.9\textwidth]{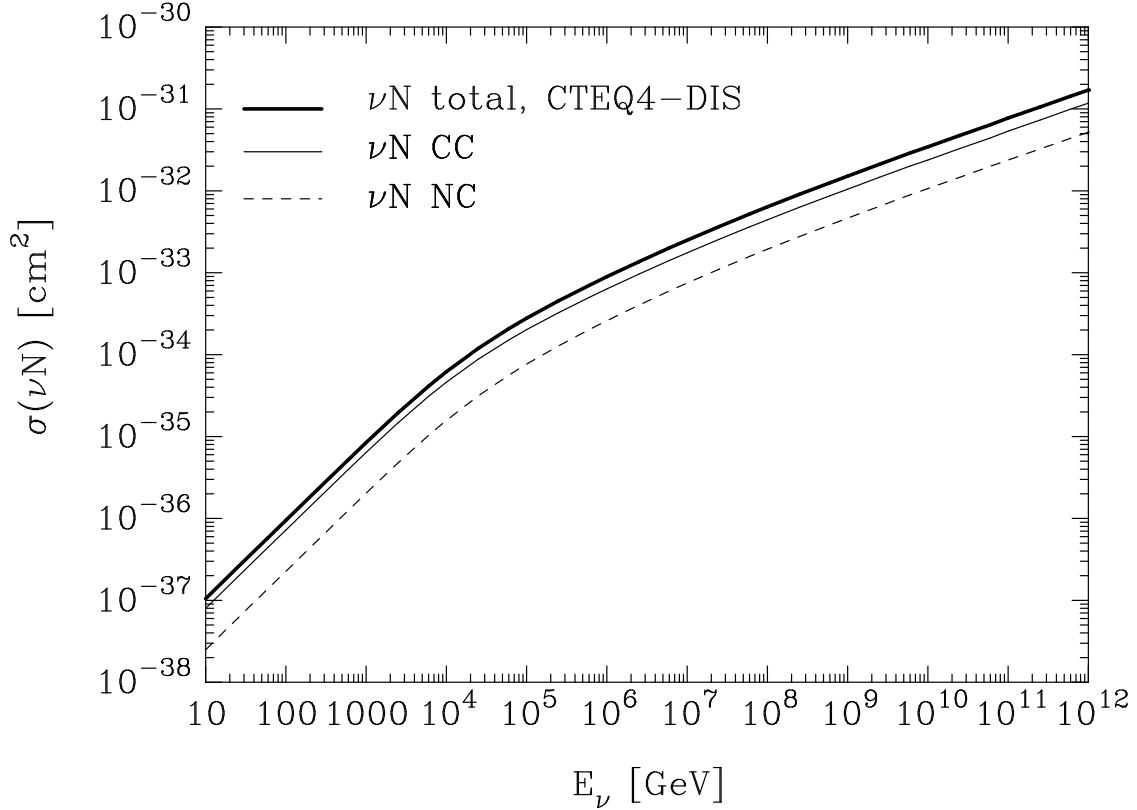}
\caption{From Ref.~\cite{Gandhi:1998ri}. Cross sections for
$\nu_{\ell} N$ interactions at high
energies, according to the CTEQ4--DIS parton distributions: dashed 
line, $\sigma(\nu_{\ell} N \rightarrow \nu_{\ell}+\hbox{anything})$; 
thin line, $\sigma(\nu_{\ell} N \rightarrow 
\ell^{-}+\hbox{anything})$; thick line, total (charged-current plus 
neutral-current) cross section.}
\label{fig1}
\end{figure}

Imagine a neutrino of energy $E_\nu$ scattering on a parton $i$
carrying a fraction $x$ of the 4-momentum $P$ of a state $X$ of mass
$M$. Denoting the fractional recoil energy of $X$ by
$y\equiv E_X^\prime/E_\nu$ and the distribution of parton type $i$ by
$f_i(x,Q)$, in the relativistic limit $E_\nu\gg m_X$ the contribution to
the $\nu X$ cross section turns out to be
\begin{equation}
  \frac{d\sigma^{\nu X}}{dxdy}=\frac{2G_{\rm F}^2ME_\nu x}{\pi}
  \left(\frac{M_{W,Z}^2}{2ME_\nu xy+M_{W,Z}^2}\right)^2
  \sum_i f_i(x,Q)\left[g_{i,L}^2+g_{i,R}^2(1-y)^2\right]\,.\label{nuX}
\end{equation}
Here, $g_{i,L}$ and $g_{i,R}$ are the left- and right-chiral couplings
of parton $i$, respectively, given by Eq.~(\ref{glr}). Eq.~(\ref{nuX})
applies to both charged and neutral currents, as well as to the
case where $X$ represents an elementary particle such as the
electron, in which case $f_i(x,Q)=\delta(x-1)$.

\begin{figure}
\includegraphics[height=0.9\textwidth,clip=true,angle=270]{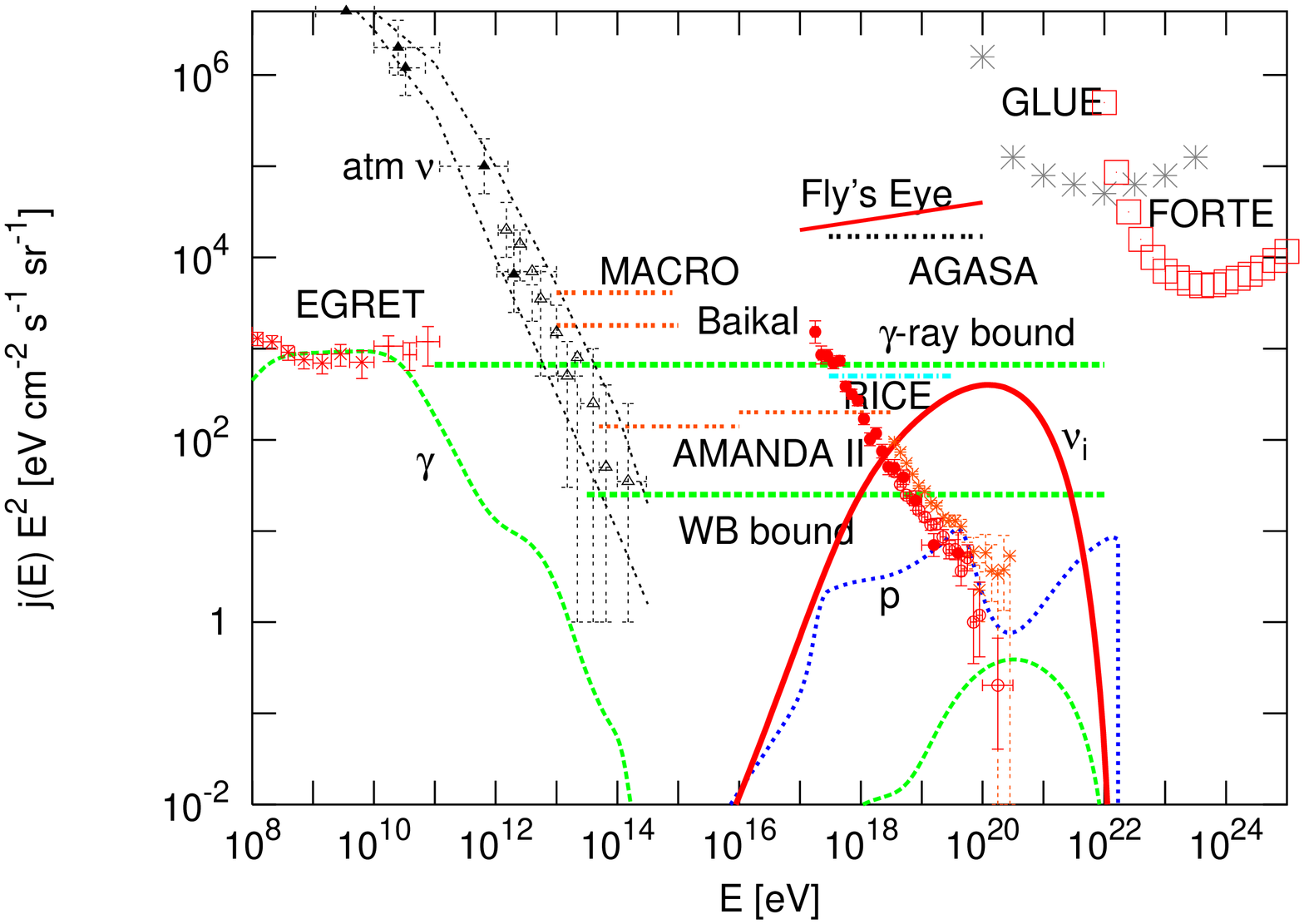}
\caption{A scenario from Ref.~\cite{ss,dima} where neutrinos are produced as
secondaries of UHE cosmic rays (data with error bars above $10^{15}\,$eV) by
interactions with the cosmic microwave background (CMB). Theoretical
fluxes of protons, $\gamma-$rays, and neutrinos (per flavor) are as indicated.
Also shown are the atmospheric neutrino flux~\cite{atm-nu}, as well as
existing upper limits on the diffuse neutrino fluxes from MACRO~\cite{MACRO},
AMANDA II~\cite{amandaII}, BAIKAL~\cite{baikal_limit},
AGASA~\cite{agasa_nu}, the Fly's Eye~\cite{baltrusaitis} and
RICE~\cite{rice_new} experiments, and the limits obtained with the
Goldstone radio telescope (GLUE)~\cite{glue} and the FORTE
satellite~\cite{forte}, as indicated. The cosmic ray data are
from the AGASA~\cite{agasa} and HiRes~\cite{hires} experiments
To the left the newest estimate of
the diffuse GeV $\gamma-$ray background from EGRET data is
shown~\cite{egret_new}.
Since most of the electromagnetic energy ends up in this energy
range and since pion production produces comparable amounts of
$\gamma-$rays and neutrinos, predicted neutrino fluxes cannot
considerably overshoot the horizontal line on the level of the
EGRET estimate marked $\gamma-$ray bound. If the sources are transparent
to produced cosmic rays, the more restrictive but less general
``Waxman-Bahcall bound''~\cite{wb-bound}, marked ``WB bound'', results.
This will be discussed in more detail in Sect.~\ref{sec_egret}.}
\label{fig2}
\end{figure}

As usual, if the four-momentum transfer $Q$ becomes comparable to
the electroweak scale, $Q^2\gg m_{W,Z}^2$, the weak gauge boson
propagator effects, represented by the factor
$M_{W,Z}^2/(Q^2+M_{W,Z}^2)$ in Eq.~(\ref{nuX}), become important.
We have used that in the limit $|Q^2|\gg M^2$ one has
$0\simeq-M^2=(xP+Q)^2\simeq Q^2+2P\cdot Qx$ with
$P\cdot Q\simeq-ME_X^\prime=-ME_\nu y$ evaluated in the laboratory
frame, i.e. the rest frame of $X$ before the interaction.
$Q^2\simeq2ME_\nu xy$ is also called the {\it virtuality}
because it is a measure for how far the exchanged gauge boson
is form the mass shell $Q^2=-M^2$.

We will not derive Eq.~(\ref{nuX}) in detail, but it is easy to
understand its structure: First, the overall normalization is
analogous to Eq.~(\ref{cross2}), using the fact that for $E_\nu\gg M$
the CM momentum $p_*\simeq M E_\nu/2$. Second, if the helicities of
the parton and the neutrino are equal, the total spin is zero
and the scattering is spherically symmetric in the CM frame.
In contrast, if the parton is right-handed, the total spin is
1 which introduces an angular dependence: After a rotation by
the scattering angle $\theta_*$ in the CM frame the particle
helicities are unchanged for the outgoing final state particle
and one has to project back onto the original helicities in order
to conserve spin. If a left-handed particle originally
propagated along the positive z-axis, its left-handed component
after scattering by $\theta_*$ in the $x-z$ plane is
\begin{equation}
  \frac{1}{2}\left(1-\frac{\ss\cdot{\bf p}}{p}\right)
  \left(\matrix{0\cr1}\right)
  =\frac{1}{2}\left(\matrix{-\sin\theta_*\cr1+\cos\theta_*}\right)\,,
\end{equation}
giving a projection $[(1+\cos\theta_*)/2]^2$. Now, Lorentz transformation
from the CM frame to the lab frame gives $E_\nu^\prime/E_\nu=
(1+v_X\cos\theta_*)/2\simeq(1+\cos\theta_*)/2$ in the relativistic
limit and thus the projection factor equals $(E_\nu^\prime/E_\nu)^2=
(1-E_X^\prime/E_\nu)^2=(1-y)^2$, as in Eq.~(\ref{nuX}) for the
right-handed parton contribution. Integrated over $0\leq y\leq1$
this gives $1/3$, corresponding to the fact that only one of
the three projections of the $J=1$ state contributes.

\subsection{Neutrino-Nucleon Scattering and Applications}\label{sec_nuN}
We now briefly consider neutrino-nucleon interaction.
From Eq.~(\ref{nuX}) it is obvious that at ultra-high energies
$2E_\nu M\gg M_{W,Z}^2$, the dominant contribution comes
from partons with
\begin{equation}
  x\sim\frac{M_{W,Z}^2}{2E_\nu M}\,.
\end{equation}
Since, very roughly, $xf_i(x,Q)\propto x^{-0.3}$ for $x\ll1$, it follows
that the neutrino-nucleon cross section grows roughly $\propto E_\nu^{0.3}$.
This is confirmed by a more detailed evaluation of Eq.~(\ref{nuX})
shown in Fig.~\ref{fig1}.

\begin{figure}
\includegraphics[height=0.9\textwidth,clip=true,angle=270]{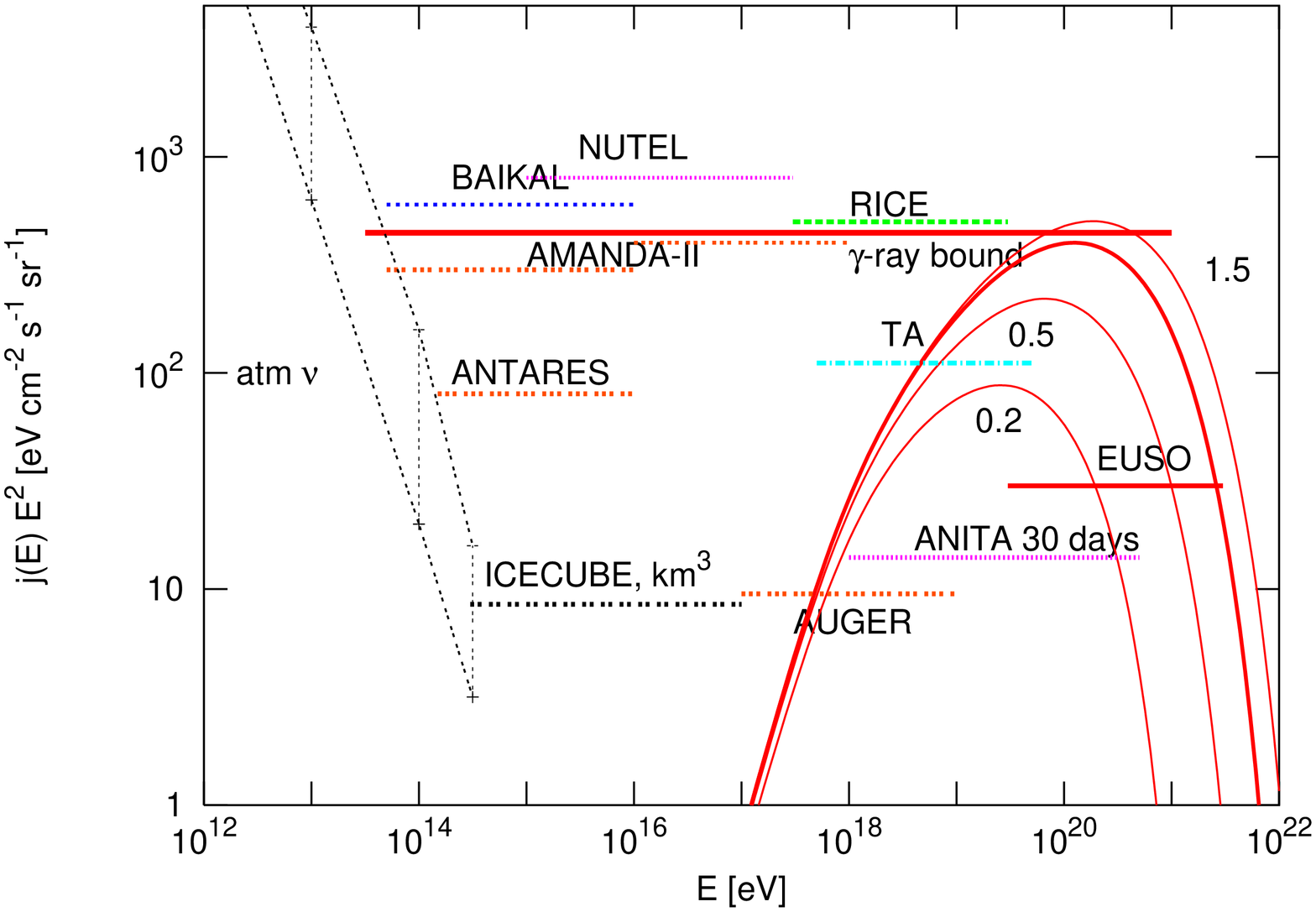}
\caption{ From Ref.~\cite{ss}. Similar to Fig.~\ref{fig2}, but showing
expected sensitivities of the currently being constructed
Pierre Auger project to tau-neutrinos~\cite{auger_nu},
the planned projects Telescope Array (TA)~\cite{ta_nu}, the
fluorescence/\v{C}erenkov detector NUTEL~\cite{mount}, the
space based EUSO~\cite{euso_nu}, the water-based Baikal~\cite{baikal_limit}
and ANTARES~\cite{antares} (the NESTOR sensitivity for 1 tower would
be similar to AMANDA-II and for 7 towers similar to ANTARES~\cite{nestor}),
the ice-based AMANDA-II~\cite{amandaII} and ICECUBE~\cite{icecube}
(similar to the intended Mediterranean km$^3$ project~\cite{katz}), and
the radio detectors RICE~\cite{rice} and ANITA~\cite{anita}, as indicated.
All sensitivities except for ANITA and RICE refer to one year running
time. For comparison, the $\gamma-$ray bound derived from the EGRET GeV
$\gamma-$ray flux~\cite{egret_new} is also shown.}
\label{fig3}
\end{figure}

Let us use this to do a very rough estimate of event rates
expected for extraterrestrial ultra-high energy (UHE) neutrinos
in neutrino telescopes.
Such neutrinos are usually produced via pion production by
accelerated UHE protons interacting within their source or
with the cosmic microwave background (CMB) during propagation to
Earth. The threshold for the reaction $N\gamma\to N\pi$, for
a head-on collision of a nucleon $N$ of energy $E$ with a photon
of energy $\varepsilon$ is given by the condition
$s=(E+\varepsilon)^2-\left[(E^2-m_N^2)^{1/2}-\varepsilon\right]^2\geq
(m_N+m_\pi)^2$, or
\begin{equation}
  E\geq\frac{m_\pi(m_N+m_\pi/2)}{2\varepsilon}\simeq
  3.4\times10^{19}\left(\frac{\varepsilon}{10^{-3}\,{\rm eV}}\right)^{-1}
  \,{\rm eV}\,,\label{gzk}
\end{equation}
where $\varepsilon\sim10^{-3}\,$eV represents the energy of a typical
CMB photon. At these energies, the secondary
neutrino flux should therefore be very roughly comparable with
the primary UHE cosmic ray flux, within large margins. Fig.~\ref{fig2}
shows a scenario where neutrinos are produced by the primary cosmic
ray interactions with the CMB. Using that the neutrino-nucleon
cross section from Fig.~\ref{fig1} roughly scales as
$\sigma_{\nu N}\propto E_\nu^{0.363}$ for
$10^{16}\,{\rm eV}\la E_\nu\la10^{21}\,$eV, and assuming
water or ice as detector medium, we obtain the rate
\begin{eqnarray}
  \Gamma_\nu&\sim&
  \sigma_{\nu N}(E_\nu)2\pi E_\nu j(E_\nu)n_N V_{\rm eff}\nonumber\\
  &\sim&0.03\left(\frac{E_\nu}{10^{19}\,{\rm eV}}\right)^{-0.637}
  \left(\frac{E_\nu^2 j(E_\nu)}
  {10^2\,{\rm eV}{\rm cm}^{-2}{\rm sr}^{-1}{\rm s}^{-1}}\right)
  \left(\frac{V_{\rm eff}}{{\rm km}^3}\right)\,{\rm yr}^{-1}\,,\label{nurate}
\end{eqnarray}
where $n_N\simeq6\times10^{23}\,{\rm cm}^{-3}$ is the nucleon
density in water/ice, $V_{\rm eff}$ the effective detection volume,
and $j(E_\nu)$ is the differential neutrino flux in units of
${\rm cm}^{-2}{\rm eV}^{-1}{\rm sr}^{-1}{\rm s}^{-1}$.

Eq.~(\ref{nurate}) indicates that at $E_\nu\ga10^{18}\,$eV,
effective volumes $\ga100\,{\rm km}^3$ are necessary. Although
impractical for conventional neutrino telescopes, big air
shower arrays such as the Pierre Auger experiment can achieve
this. In contrast, if there are sources such as active galactic
nuclei emitting at $E_\nu\sim10^{16}\,$eV at a level
$E_\nu^2 j(E_\nu)\sim10^2\,{\rm eV}{\rm cm}^{-2}{\rm sr}^{-1}{\rm s}^{-1}$,
km-scale neutrino telescopes should detect something. Such fluxes
are consistent with general considerations, see Fig.~\ref{fig2}.

Finally, Fig.~\ref{fig3} shows more detailed neutrino flux
sensitivities expected from future experiments.

\section{Dirac and Majorana Neutrinos}\label{sec_dirana}
Up to now we have assumed that neutrinos and anti-neutrinos are
separate entities. This is true if lepton number is conserved,
see Tab.~\ref{tab1}, and corresponds to pure "Dirac neutrinos".
However, lepton number may be violated in the neutrino sector
and neutrinos may be indistinguishable from anti-neutrinos.
In order to elucidate this, let us first study some symmetries
of the Dirac equation~(\ref{dirac}). From Eqs.~(\ref{clifford}),
(\ref{gammamunu}) one can easily show that
\begin{equation}
  \gamma_\mu^*=\gamma_2\gamma_\mu\gamma_2\,.\label{complexconj}
\end{equation}
Complex conjugating the Dirac equation (\ref{dirac}) and multiplying
it with $\xi^*\gamma_2$ from the left, it then follows that it is
invariant under the "charge conjugation transformation"
\begin{equation}
  C\psi(x)C^{-1}\equiv\psi^c\equiv\xi^*\gamma_2\psi^*
  \,,\label{chargeconj}
\end{equation}
where $\xi$ is an arbitrary complex number with $|\xi|=1$. This
transformation exchanges particles and anti-particles and satisfies
$(\psi^c)^c=\psi$. Note that $\gamma_2$ appears because according
to Eq.~(\ref{complexconj}) it is the only real Dirac matrix in this
convention.

A {\sl Majorana neutrino} satisfies the reality condition
\begin{equation}
  \phi(x)=\gamma_2\phi^*(x)\,,\quad\mbox{or}\quad\phi^c=\xi^*\phi\,.
  \label{majorana}
\end{equation}
These spinors are of the form
\begin{equation}
  \phi=\left(\matrix{-i\sigma_2\chi^*\cr\chi}\right)\,,
  \label{majorana2}
\end{equation}
where $\chi$ is a two-spinor. This implies that for any Dirac spinor
$\psi$ one can construct a Majorana spinor by
\begin{equation}
  \phi\equiv\psi+\xi\psi^c\,.
\end{equation}
Note that this spinor is not an eigenstate of lepton number
because under a phase transformation $\psi\to\psi e^{i\alpha}$
one has $\psi^c\to\psi^c e^{-i\alpha}$. Defining
\begin{equation}
  \psi^c_{L,R}\equiv(\psi^c)_{L,R}=\frac{1\pm\gamma_5}{2}\psi^c=
  \left(\frac{1\mp\gamma_5}{2}\psi\right)^c\,,
\end{equation}
one can define left- and right-handed Majorana fields
\begin{equation}
  \phi_\pm\equiv\psi_{L,R}+(\psi_{L,R})^c=
  \psi_{L,R}+\psi_{R,L}^c\,.\label{phipm}
\end{equation}
Note that both these fields now contain both left and right-handed
fields. What before experimentally was called neutrino and anti-neutrino
now is called left- and right-handed neutrino, respectively.
We can now introduce Majorana mass terms of the form
\begin{eqnarray}
  {\cal L}_{\rm M}&=&-\frac{1}{2}\left(m_L\overline{\phi_+}\phi_+
  +m_R\overline{\phi_-}\phi_-\right)\label{majoranamass}\\
  &=&-\frac{1}{2}m_L\left(\overline{\psi_L}\psi_R^c+\overline{\psi_R^c}\psi_L
  \right)
  -\frac{1}{2}m_R\left(\overline{\psi_R}\psi_L^c+\overline{\psi_L^c}\psi_R
  \right)\,,\nonumber
\end{eqnarray}
where $m_L$ and $m_R$ are real.
Together with the Dirac mass term this can be written as
\begin{equation}
  {\cal L}_{\rm M}+{\cal L}_{\rm D}=-\frac{1}{2}
  \left(\overline{\psi_L},\overline{\psi_L^c}\right)
  \left(\matrix{m_L&m_{\rm D}\cr m_{\rm D}&m_R}\right)
  \left(\matrix{\psi_R^c\cr\psi_R}\right)+{\rm h.c.}\,,\label{mdmass}
\end{equation}
where we have used $(\overline{\psi_1}\psi_2)^\dagger=
\overline{\psi_2}\psi_1$, see around Tab.~\ref{tab2} and
(using $\gamma_2^\dagger=\gamma_2$ and
the fact that $\psi$ anti-commutes)
\begin{eqnarray}
  \overline{\psi_L^c}\psi_R^c&=&\left[(\psi_R)^c\right]^\dagger
  i\gamma^0\frac{1-\gamma_5}{2}\xi^*\gamma_2\psi^*=
  \psi_R^T\gamma_2^\dagger i\gamma^0\frac{1-\gamma_5}{2}\gamma_2\psi^*
  \nonumber\\
  &=&-i\psi_R^T\frac{1-\gamma_5}{2}\gamma^0\psi^*=
  i\left[\psi_R^T\frac{1-\gamma_5}{2}\gamma^0\psi^*\right]^T=
  \overline{\psi_L}\psi_R\nonumber
\end{eqnarray}
for any Dirac spinors $\psi_{1,2}$ and $\psi$. Note that under
$\psi\to\psi e^{i\alpha}$ Dirac terms are invariant, whereas
Majorana terms pick up the phase $e^{\pm2i\alpha}$, according to
lepton number conservation and non-conservation, respectively.
Furthermore, we see that for $m_R\gg m_{\rm D}$, $m_L\simeq0$, the
two mass eigenvalues in Eq.~(\ref{mdmass}) are $\simeq m_R$ and
$m_{\rm D}^2/m_R$. The latter are very small and thus may explain
the sub-eV masses involved in left-chiral neutrino oscillations.
This is called the {\it see-saw mechanism} which would imply that the
mass eigenstates are Majorana in nature. The existence of one heavy
right-handed Majorana neutrino per lepton generation is motivated
by Grand Unification extensions of the electroweak gauge group
to $SO(10)$ which has 16-dimensional representations that could
fit 15 Standard Model lepton and quark states plus one new state,
see, e.g., Ref.~\cite{mohapatra}.

Finally, in the exactly massless case, Dirac and Majorana particles
are exactly equivalent, since the two fields Eq.~(\ref{phipm})
completely decouple, see Eq.~(\ref{mdmass}). We also mention
that in supersymmetric extensions of the Standard Model, the
fermionic super-partners of the gauge bosons are Majorana fermions.
As a consequence, they can self-annihilate which plays an important
role to their being candidates for {\it cold dark matter}.

For $n>1$ neutrino flavors, mass eigenstates $\left|\nu_i\right>$
of mass $m_i$ and interaction eigenstates $\left|\nu_\alpha\right>$
in general are not identical, but related by a unitary $n\times n$
matrix $U$:
\begin{equation}
  \left|\nu_\alpha\right>=\sum_i U_{\alpha i}\left|\nu_i\right>\,,
  \label{U}
\end{equation}
where for anti-neutrinos $U$ has to be replaced by $U^*$. Such
a matrix in general has $n^2$ real parameters. Subtracting
$2n-1$ relative phases of the $n$ neutrinos in the two bases,
one ends up with $(n-1)^2$ physically independent real parameters.
Of these, $n(n-1)/2$ are mixing angles, and the remaining
$(n-1)(n-2)/2$ are "$CP$-violating phases". In order to have $CP$-violation
in the Dirac neutrino sector thus requires $n\geq3$. Once the relative
phases of the different flavors have been fixed, for non-vanishing
Majorana masses there will in general be $n-1$ Majorana phases
that can not be projected out by $\psi\to\psi e^{i\alpha}$ in
Eq.~(\ref{majoranamass}). Thus, the number of independent real
parameters is larger, namely $n(n-1)$, in this case. Note that the
corresponding {\it Cabibbo Kobayashi Maskawa (CKM)} matrix
in the quark sector is pure Dirac because Majorana terms
would violate electric charge conservation in the quark sector.

If at time $t=0$ a flavor eigenstate $\left|\nu_\alpha\right>=
\sum_i U_{\alpha i}\left|\nu_i\right>$ is produced in an interaction,
in vacuum the time development will thus be
\begin{equation}
  \left|\nu(t)\right>=\sum_i U_{\alpha i}e^{-iE_it}\left|\nu_i\right>=
  \sum_{i,\beta} U_{\alpha i}U_{\beta i}^*e^{-iE_it}\left|\nu_\beta\right>\,.
  \label{nuosc}
\end{equation}
Since masses and energies of anti-particles are equal according
to the $CPT$ theorem, from this we obtain the following transition
probabilities
\begin{eqnarray}
  P(\nu_\alpha\to\nu_\beta)&=&\left|\sum_i U_{\alpha i}U^*_{\beta i}
  \exp(-iE_it)\right|^2\nonumber\\
  P(\bar\nu_\alpha\to\bar\nu_\beta)&=&\left|\sum_i U^*_{\alpha i}U_{\beta i}
  \exp(-iE_it)\right|^2\label{tranprob}\,.
\end{eqnarray}
From this follows immediately
\begin{equation}
  P(\nu_\alpha\to\nu_\beta)=P(\bar\nu_\beta\to\bar\nu_\alpha)\,,
\end{equation}
which is due to the $CPT$ theorem. Furthermore, if the mixing matrix
satisfies a reality condition of the form
\begin{equation}
  U_{\alpha i}=U^*_{\alpha i}\eta_i\,,
\end{equation}
with $\eta_i$ phases, corresponding to $CP$-conservation, one also
has
\begin{equation}
  P(\nu_\alpha\to\nu_\beta)=P(\bar\nu_\alpha\to\bar\nu_\beta)\,.
\end{equation}

We mention two other important differences between Dirac and Majorana
neutrinos:

\begin{itemize}
\item Neutrino-less double beta-decay is only possible in the presence
of Majorana masses because the final state $e^-e^-$ violates lepton
number. In this case the rate is proportional to the square of
\begin{equation}
  m_{ee}=\left|\sum_i|U_{ei}|^2m_ie^{i\alpha_i}\right|\,,
\end{equation}
see Eq.~(\ref{mdmass}),
where only one of the Majorana phases $\alpha_i$ can be projected out.
Apart from these phases, this equation results from Eq.~(\ref{nuosc})
for $\alpha=\beta=e$. There is even evidence claimed for this kind of decay,
and thus for an electron neutrino Majorana mass around 0.4 eV,
see Ref.~\cite{klapdor}. The issue is expected to be settled by
next generation experiments such as CUORE~\cite{cuore}.

In contrast, in $\beta-$decay with neutrinos, the electron spectra
are influenced by the individual eigenstates of real mass $m_i$, and
not by any phases. The current best experimental upper limit
is given by the Mainz experiment based on tritium $\beta-$decay
$^3{\rm H}\to^3{\rm He}e^-\bar\nu_e$~\cite{mainz},
\begin{equation}
  m_{\nu_e}=\sqrt{\sum_i|U_{ei}|^2m^2_i}\la 2.2\,{\rm eV}\label{nudirect}
\end{equation}
at 95\% confidence level (CL).
The KATRIN experiment~\cite{katrin} aims at a sensitivity down to $0.2\,$eV
within the next few years.

\item Majorana neutrinos cannot have magnetic dipole moments
between equal neutrino flavors, as seen from the following identity
using Eqs.~(\ref{majorana}), (\ref{complexconj}), (\ref{dagger}), and
the reality of the spinors in Tab.~\ref{tab2}:
\begin{eqnarray}
  &&i\overline{\psi_1}[\gamma_\mu,\gamma_\nu]\psi_2=
  -\psi_1^T\gamma_2\gamma^0[\gamma_\mu,\gamma_\nu]\gamma_2\psi_2^*=
  -\psi_1^T\left(\gamma^0[\gamma_\mu,\gamma_\nu]\right)^*\psi_2^*=\nonumber\\
  &&=-\left(\psi_1^\dagger\gamma^0[\gamma_\mu,\gamma_\nu]\psi_2\right)^*=
  \left(\psi_1^\dagger\gamma^0[\gamma_\mu,\gamma_\nu]\psi_2\right)^\dagger=
  -i\overline{\psi_2}[\gamma_\mu,\gamma_\nu]\psi_1\,,\nonumber
\end{eqnarray}
where in the second-last identity we have used that transposition changes
the order of the fermionic fields, thus picking up a minus sign. As a
consequence, only {\it transition magnetic moments} between different
flavors are possible for Majorana neutrinos.
\end{itemize}

\section{Neutrino Oscillations}
Let us now restrict to two-neutrino oscillations, $n=2$, between
$\left|\nu_e\right>$ and $\left|\nu_\mu\right>$, say, and write
\begin{equation}
  U=\left(\matrix{\cos\theta_0&
  \sin\theta_0\cr-\sin\theta_0&\cos\theta_0}\right)\label{U2}
\end{equation}
for the {\it mixing matrix} in Eq.~(\ref{U}) which is characterized
by one real vacuum mixing angle $\theta_0$. Since
$id\left|\nu_i\right>/dt=E_i\left|\nu_i\right>$ for $i=1,2$ in the
mass basis, and since $E_i=(m_i^2+p^2)^{1/2}\simeq p+m_i^2/(2p)\simeq
E+m_i^2/(2E)$
in the relativistic limit $p\gg m_i$, using the trigonometric identities
$\cos^2\theta_0-\sin^2\theta_0=\cos2\theta_0$,
$2\cos\theta_0\sin\theta_0=\sin2\theta_0$, it follows from Eqs.~(\ref{U}),
(\ref{U2}) that
\begin{equation}
  i\frac{d}{dt}\left(\matrix{\nu_e\cr\nu_\mu}\right)_E=
  \left[\left(E+\frac{m_1^2+m_2^2}{4E}\right)+
  \frac{\Delta m^2}{4E}
  \left(\matrix{\cos2\theta_0&-\sin2\theta_0\cr-\sin2\theta_0&-\cos2\theta_0}
  \right)\right]\left(\matrix{\nu_e\cr\nu_\mu}\right)_E\,,\label{oscvac}
\end{equation}
where we consider a given momentum mode ${\bf p}$, and
$\Delta m^2\equiv m_1^2-m_2^2$. From now on we will consider the
relativistic limit with $p\simeq E$.
The first term in Eq.~(\ref{oscvac}) is a common phase factor and
can be ignored.

The integrated version of this is Eq.~(\ref{nuosc}). Then applying
Eq.~(\ref{tranprob}), one can show that this has the solution
\begin{equation}
  P(\nu_\alpha\to\nu_\beta)=\frac{1}{2}\sin^22\theta_0
  \left(1-\cos\Delta m^2\frac{L}{2E}\right)\,\quad\mbox{for}
  \quad\alpha\neq\beta\,,\label{nuosc2}
\end{equation}
for oscillations over a length $L$. The oscillation length in
vacuum is thus
\begin{equation}
  L_0=4\pi\frac{E}{|\Delta m^2|}\simeq2.48\,\left(\frac{E}{{\rm MeV}}\right)
  \left(\frac{|\Delta m^2|}{{\rm eV}^2}\right)^{-1}\,{\rm m}
  \,.\label{osclength}
\end{equation}

Neutrino oscillations are modified by forward scattering amplitudes
in matter. Since neutral currents are by definition flavor-neutral,
they only contribute to the common phase factor which in the following
will be ignored. The charged current interaction is diagonal in flavor
space and, according to Eqs.~(\ref{ccnc}),~(\ref{ccnc2}), and~(\ref{gf}),
the low-energy limit of its forward scattering part for $\nu_e({\bf p})$
has the form
\begin{equation}
  2\sqrt2G_{\rm F}\sum_{{\bf p}^\prime}
  \left[e({\bf p}^\prime)^\dagger\gamma^0
  \gamma^\mu\frac{1+\gamma_5}{2}\nu_e({\bf p})\right]
  \left[\nu_e({\bf p})^\dagger\gamma^0
  \gamma_\mu\frac{1+\gamma_5}{2}e({\bf p}^\prime)\right]\,,\label{msw}
\end{equation}
where we have used $\bar\psi=\psi^\dagger i\gamma^0$. We need to
express this in terms of Dirac bilinears of the form Tab.~\ref{tab2}
for electrons and neutrinos separately. In order to do that we
use the fact that every $4\times4$ matrix ${\cal O}$ can be expanded
according to
\begin{equation}
  {\cal O}=\sum_i\frac{{\rm tr}({\cal O}{\cal O}_i)}{{\rm tr}({\cal O}_i^2)}
  {\cal O}_i\,,
\end{equation}
where ${\cal O}_i$ are the 16 matrices appearing in Tab.~\ref{tab2}
which satisfy ${\rm tr}({\cal O}_i{\cal O}_j)=0$ for $i\neq j$.
Using this one can show that
\begin{eqnarray}
  &&\left(\gamma^0\gamma^\mu\frac{1+\gamma_5}{2}\right)_{\alpha\beta}
  \left(\gamma^0\gamma_\mu\frac{1+\gamma_5}{2}\right)_{\gamma\delta}=
  \nonumber\\
  &&\left(\frac{1+\gamma_5}{2}\right)_{\gamma\beta}
  \left(\frac{1+\gamma_5}{2}\right)_{\alpha\delta}+
  \frac{1}{8}\left(\left[\gamma^\lambda,\gamma^\kappa\right]
  \frac{1+\gamma_5}{2}\right)_{\gamma\beta}
  \left(\left[\gamma_\lambda,\gamma_\kappa\right]\right)_{\alpha\delta}
  \,.\label{diracexp}
\end{eqnarray}
Applying this to Eq.~(\ref{msw}) and noting that the last term in
Eq.~(\ref{diracexp}) does not contribute in the rest frame of the
electron plasma, one obtains
\begin{equation}
  -2\sqrt2G_{\rm F}
  \left[\nu_e({\bf p})^\dagger\frac{1+\gamma_5}{2}\nu_e({\bf p})\right]
  \sum_{{\bf p}^\prime}
  \left[e({\bf p}^\prime)^\dagger\frac{1+\gamma_5}{2}e({\bf p}^\prime)\right]
  \,,\label{msw2}
\end{equation}
where we have picked up an extra minus from the anti-commutation of
fermionic fields. The electron and neutrino fields have the form
Eq.~(\ref{expansion}). The matter-dependent part of the sum in
Eq.~(\ref{msw2}) thus takes the form $\sum_{\bf p}(a_{e,L}^\dagger({\bf p})
a_{e,L}({\bf p})-b_{e,L}^\dagger({\bf p}) b_{e,L}({\bf p})$. When tracing
out the charged lepton density matrix, this reduces to the density of
left-chiral electrons minus the density of left-chiral positrons.
Thus, for an unpolarized plasma, the contribution to the
$\left|\nu_e\right>$ self-energy finally is $-\sqrt2G_{\rm F}N_e$,
where $N_e$ is the electron-number
density, i.e. the electron minus the positron density, and analogously
for the other active flavors. The non-trivial
part of Eq.~(\ref{oscvac}) is thus modified to
\begin{equation}
  i\frac{d}{dt}\left(\matrix{\nu_e\cr\nu_\mu}\right)_E=
  \left(\matrix{\frac{\Delta m^2\cos2\theta_0}{4E}-\sqrt2G_{\rm F}N_e&
  -\frac{\Delta m^2\sin2\theta_0}{4E}\cr-\frac{\Delta m^2\sin2\theta_0}{4E}&
  -\frac{\Delta m^2\cos2\theta_0}{4E}-\sqrt2G_{\rm F}N_\mu}\right)
  \left(\matrix{\nu_e\cr\nu_\mu}\right)_E\,.\label{oscmat}
\end{equation}

It is illustrative to write this in terms of the hermitian density matrix
$\rho_{\bf p}(t)\equiv\left.|\nu_{\bf p}(t)\right\rangle\left\langle\nu_{\bf p}(t)\right.|$.
If we expand this into an occupation number $n_{\bf p}$ and polarization
${\bf P_p}$, $\rho_{\bf p}=\frac{1}{2}\left(n_{\bf p}+
{\bf P_p}\cdot\ss\right)$, one can easily show that Eq.~(\ref{oscmat})
is equivalent to
\begin{equation}
  \dot{\bf P}_{\bf p}={\bf B_p}\times{\bf P_p}\,,\label{denseq}
\end{equation}
where the precession vector ${\bf B_p}\equiv{\bf B}({\bf p})$
\begin{eqnarray}
  B_1({\bf p})&=&\frac{\Delta m^2}{2E}\sin2\theta_0\,,\nonumber\\
  B_3({\bf p})&=&\frac{\Delta m^2}{2E}\cos2\theta_0-\sqrt2 G_{\rm F}N\,,
  \label{precession}
\end{eqnarray}
where $N=N_e-N_{\mu,\tau}$ for $\nu_e-\nu_{\mu,\tau}$ mixing and
$N=N_{e,\mu,\tau}-N_n$ for active-sterile mixing, with $N_n$ a
combination of nucleon densities.
 Thus, neutrino oscillations are
mathematically equivalent to the precession of a magnetic moment
in a variable external magnetic field.

Eq.~\ref{precession}) shows immediately that at the {\it resonance density}
\begin{equation}
  N_r\equiv\frac{\Delta m^2\cos2\theta_0}{2\sqrt2 G_{\rm F}E}
\end{equation}
the two diagonal entries in Eq.~(\ref{oscmat}) become equal or,
equivalently, $B_3({\bf p})=0$. More generally,
Eqs.~(\ref{oscmat}), (\ref{denseq}), (\ref{precession}) are diagonalized
by a mixing matrix Eq.~(\ref{U2}) where $\theta_0$ is replaced by
the mixing angle in matter $\theta_E$ given by
\begin{equation}
  \tan2\theta_E=\frac{\tan2\theta_0}{1-N/N_r}\,.
\end{equation}
Maximum mixing, $\theta_E=\pi/4$, thus occurs at the so-called
{\it Michaev-Smirnow-Wolfenstein (MSW) resonance} at $N=N_r$.
For $N\ll N_r$ one thus has vacuum mixing, $\theta_E\simeq\theta_0$,
whereas for $N\gg N_r$ one has $\theta_E\ll\theta_0$. Eq.~(\ref{denseq})
now shows that propagation from $N\gg N_r$ to $N\ll N_r$ can lead to
an efficient transition from one flavor to another, as long as
$|\dot{\bf B}_{\bf p}|\la|{\bf B_p}|^2$. Such transitions
are called {\it adiabatic}. Note that since masses of anti-particles
and particles are
equal, whereas the lepton numbers $N_{e,\mu,\tau}$ change sign
under charge conjugation, resonances in matter occur either for
neutrinos or for anti-neutrinos.

In subsequent sections we will apply this to various observational
evidence for neutrino oscillations.

\section{Selected Applications in Astrophysics and Cosmology}
\subsection{Stellar Burning and Solar Neutrino Oscillations}\label{sec_sol}
Weak interactions are crucial in cosmology and stellar physics.
In main sequence stars the first stage of hydrogen fusion into
helium is the weak interaction
\begin{equation}
  p+p\to\mbox{$^2$H}+e^++\nu_e\,.
\end{equation}
The subsequent reactions $^2$H$+p\to^3$He$+\gamma$ and
$^3$He$+^3$He$\to^4$He$+p+p$ lead to the net reaction
\begin{equation}
  4p\to\mbox{$^4$He}+2e^++2\nu_e+26.73\,{\rm MeV}\,.
\end{equation}
For many more details see Ref.~\cite{raffelt}.

\begin{figure}
\includegraphics [width=0.8\textwidth,angle=270]{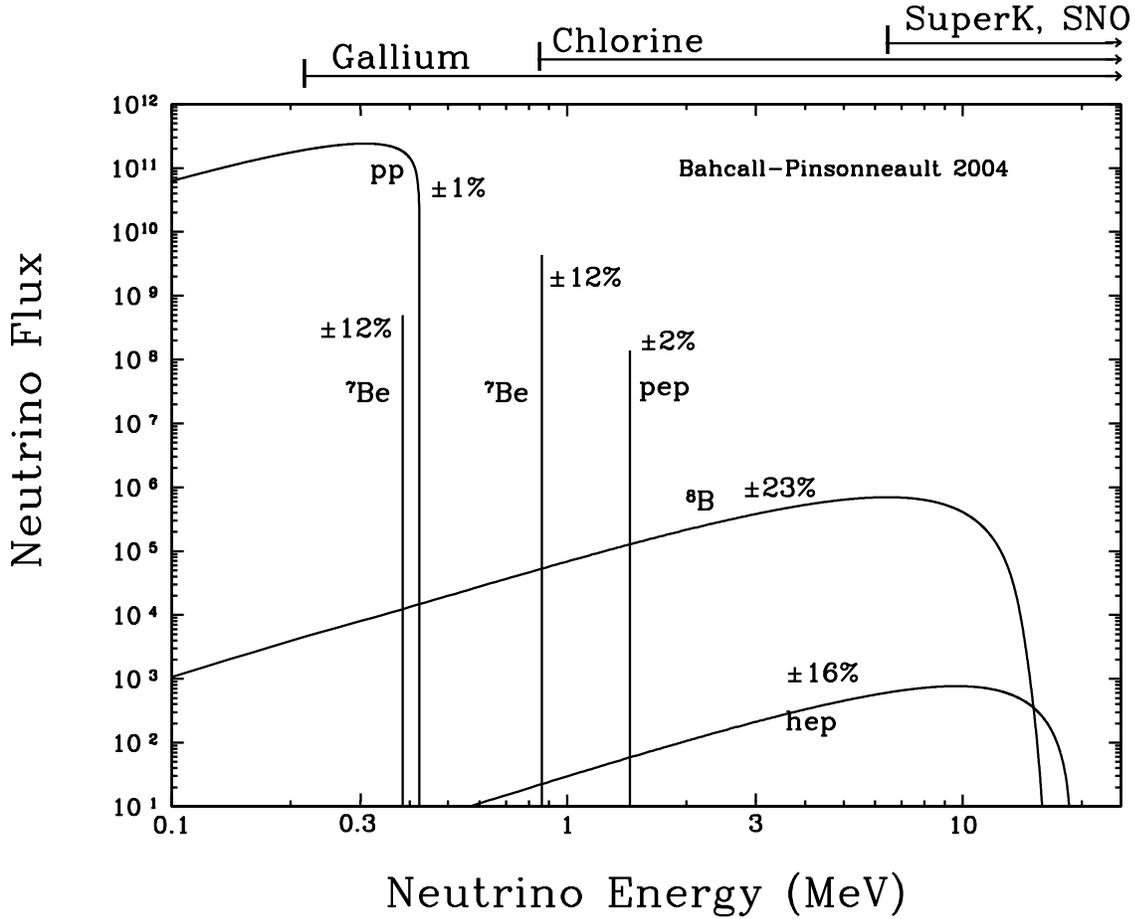}
\caption{The predicted solar neutrino energy spectrum from Ref.~\cite{bp04}.
For continuum sources, the neutrino
fluxes are given in number per ${\rm cm^{-2} sec^{-1} MeV^{-1}}$
at the Earth's surface. For line sources, the units are number per
${\rm cm^{-2} sec^{-1}}$. The total theoretical uncertainties
are shown. The CNO neutrino fluxes are not shown.\label{fig4}}
\end{figure}

When normalized to the solar energy flux arriving at Earth,
one can calculate the expected neutrino fluxes within the
so-called Standard Solar Model. The resulting fluxes in this ``pp'' channel
as well as in other reaction channels are shown in Fig.~\ref{fig4}

\begin{figure}
\includegraphics [width=0.9\textwidth]{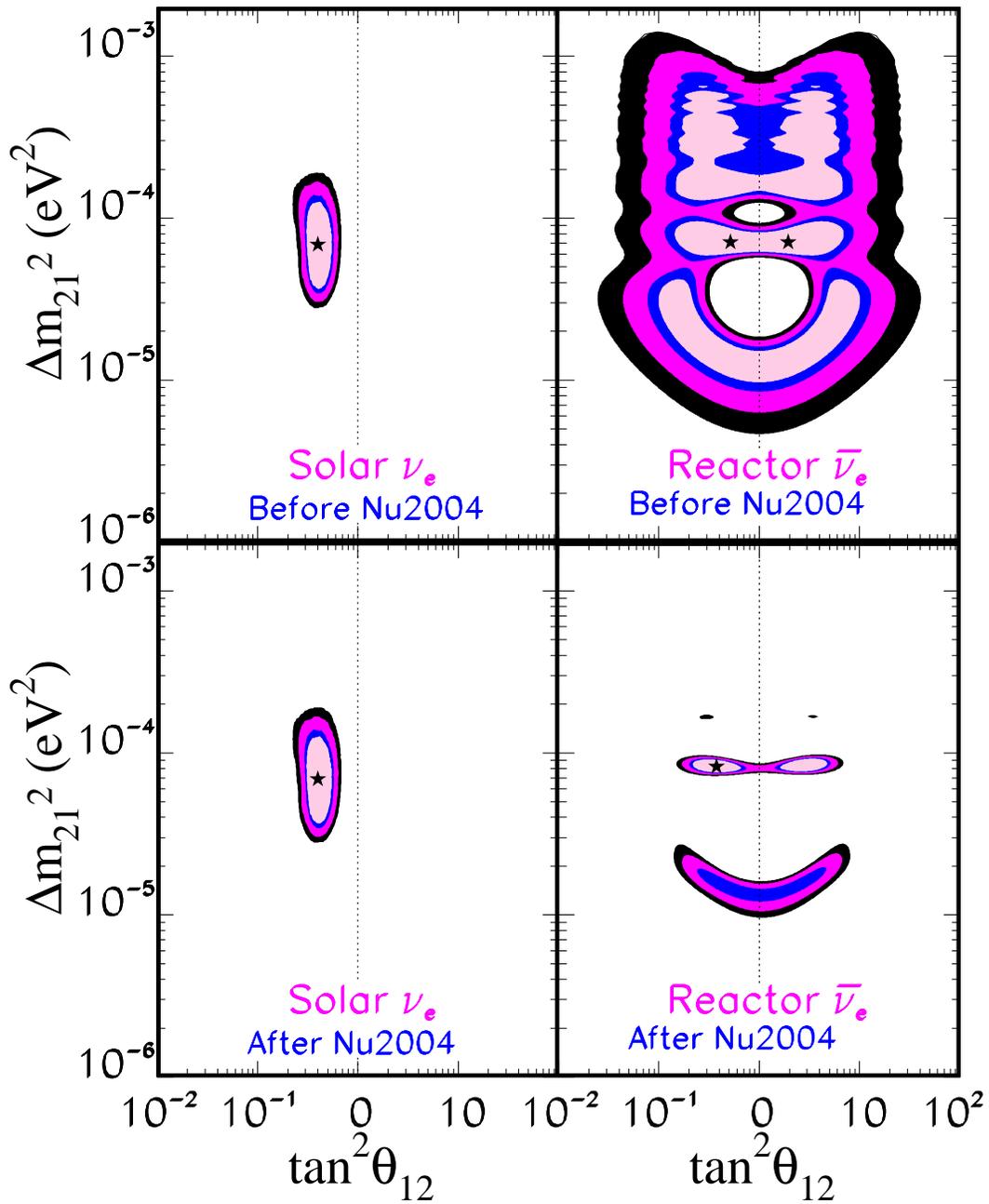}
\caption{From Ref.~\cite{bahcall}. Allowed oscillation parameters:
Solar vs KamLAND.
The two left panels show the 90\%, 95\%, 99\%, and 3$\sigma$
allowed regions for oscillation parameters that are obtained by a
global fit of all the available solar data.
The two right panels show the 90\%, 95\%, 99\%, and 3$\sigma$
allowed regions for oscillation parameters that are obtained by a
global fit of all the reactor data from KamLAND and
CHOOZ. The two upper (lower) panels
correspond to the analysis of all data available before (after)
the Neutrino 2004 conference, June 14-19, 2004 (Paris). The new
KamLAND data are sufficiently precise that
matter effects discernibly break the degeneracy between the two
mirror vacuum solutions in the lower right panel.
\label{fig5}}
\end{figure}

However, less than half of the expected solar electron-neutrino flux
at a few MeV has been observed. On the other hand, neutral current
experiments with the Sudbury Neutrino Observatory (SNO)~\cite{sno} have
shown that the sum of the electron, muon- and
tau neutrino flux coincides with the expected electron neutrino
flux. This can be explained by an MSW transition of $\nu_e$ into
$\nu_\mu$ and $\nu_\tau$ within the Sun with a
$\Delta m^2_{\rm solar}\sim10^{-5}\,{\rm eV}^2$.
Note that this corresponds to a vacuum oscillation length
Eq.~(\ref{osclength}) of a few hundred kilometers.
Recently this has been confirmed independently by the KamLAND
experiment~\cite{kamland} which
measured the disappearance of the $\bar\nu_e$ neutrinos produced
by nuclear reactors a few hundred kilometers from the detector.
The best fit parameters for the parameters of mixing of two neutrinos
in vacuum from all solar and reactor data are~\cite{bahcall}
\begin{equation}
  \Delta m^2_{\rm solar}\simeq8.2^{+0.3}_{-0.3}(^{+1.0}_{-0.8})
  \times10^{-5}\,{\rm eV}^2\,;\quad
  \tan^2\theta_{\rm solar}\simeq0.39^{+0.05}_{-0.04}(^{+0.19}_{-0.11})
  \,,\label{solar}
\end{equation}
where $1\sigma$ and $3\sigma$ errors are given.
The relevant contour plots are shown in Fig.~\ref{fig5}.
It is interesting to note that maximal mixing is strongly excluded,
$\tan\theta_{\rm solar}\leq1.0$ at $5.8\sigma$.

Finally, since solar and reactor neutrino experiments deal
with $\nu_e$ and $\bar\nu_e$, respectively, comparison of
the two corresponding oscillations parameters allows to
set limits on $CPT$-violation in the neutrino sector.

\subsection{Atmospheric Neutrinos}\label{sec_atm}
\begin{figure}
\includegraphics[width=0.9\textwidth]{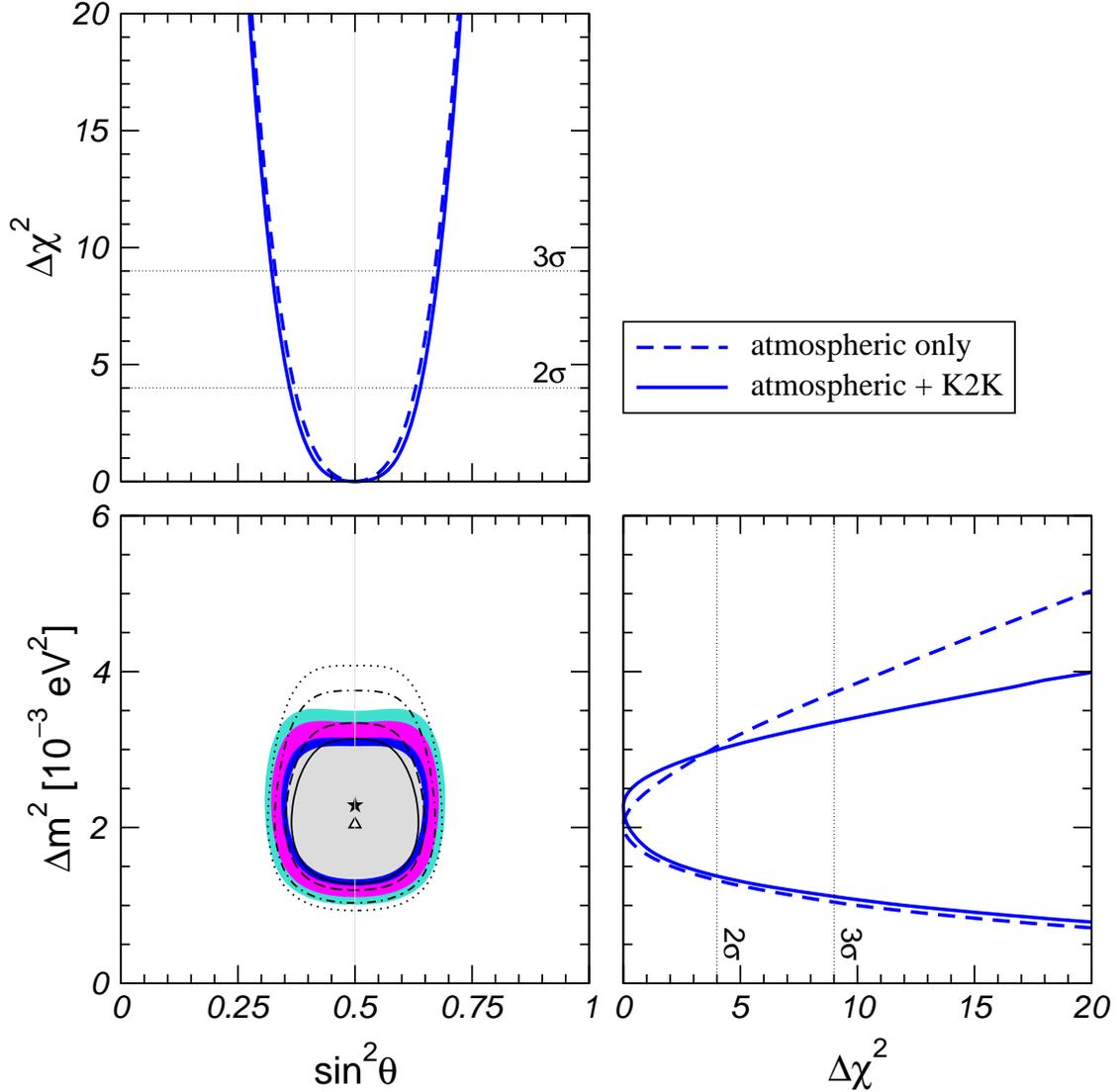}
\caption{From Ref.~\cite{maltoni}.
Allowed ($\sin^2\theta_{\rm atmos}$,~$\Delta m^2_{\rm atmos}$)
regions at 90\%, 95\%, 99\%, and 3$\sigma$ CL for two
degrees of freedom. The regions
delimited by the lines correspond to atmospheric data only, while for the
colored regions also K2K data are added. The best fit point of
atmospheric (atmospheric + K2K) data is marked by a triangle
(star). Also shown is the $\Delta \chi^2$ as a function of
$\sin^2\theta_{\rm atmos}$ and $\Delta m^2_{\rm atmos}$,
minimized with respect to the
undisplayed parameter.\label{fig6}}
\end{figure}

Cosmic rays interact in the atmosphere and produce, among
other particles, pions and kaons whose decay products
contain neutrinos.
The observed ratio of upcoming to down-going atmospheric muon
neutrinos is about 0.5 in the GeV range. Since upcoming neutrinos
travel several thousand kilometers, this can be interpreted as
vacuum oscillations between muon and tau-neutrinos. The best fit
parameters are
\begin{equation}
  \Delta m^2_{\rm atmos}\simeq2.1\times10^{-3}\,{\rm eV}^2\,;\quad
  \sin^2\theta_{\rm atmos}\simeq0.5\,,\label{atmospheric}
\end{equation}
consistent with maximal mixing. The corresponding contours
resulting from atmospheric and long baseline neutrino oscillation
data from K2K~\cite{k2k} are shown in Fig.~\ref{fig6}. Recently, the
$L/E$ dependence, see Eq.~(\ref{nuosc2}), characteristic for neutrino
oscillations has been confirmed by the Superkamiokande
experiment~\cite{superk},
thereby strongly constraining alternative explanations of
muon neutrino disappearance such as neutrino decay~\cite{ashie}.
In addition, oscillations into sterile neutrinos are strongly
disfavored over oscillations into tau-neutrinos via the following
discriminating effects: Neutral currents would be non-diagonal
for oscillations into sterile states, thus modifying oscillation
amplitudes and total scattering rates, and charged current
interactions of tau-neutrinos imply $\tau$ appearance.

In 3-neutrino oscillation schemes $\theta_{\rm solar}$ and
$\theta_{\rm atmos}$ are usually identified with $\theta_{12}$
and $\theta_{23}$, respectively  (we here assume the ``normal mass
hierarchy'' $m_1,m_2\ll m_3$ which seems the most natural for
neutrino mass modeling in Grand Unification scenarios~\cite{altarelli}).
According to the discussion around
Eq.~(\ref{U}), there is one more mixing angle $\theta_{13}$ and
at least one Dirac $CP$-violating phase called $\delta$. Note that
solar and atmospheric neutrinos only decouple exactly for
$\theta_{13}=0$ in which case $CP$ would also be conserved.
Whereas there are at most weak indications for leptonic
$CP$-violations yet~\cite{klinkhamer}, the third mixing angle is
constrained at 3$\sigma$ CL by Ref.~\cite{maltoni}
\begin{equation}
  \sin^2\theta_{13}\leq0.061\,.
\end{equation}

We finally stress that neutrino oscillations are sensitive
only to differences of squared masses, not to absolute mass
scales. To probe the latter requires laboratory experiments
discussed earlier such as $\beta-$decay, the study of cosmological
effects such as the influence of neutrino mass on the power spectrum,
see Sect.~\ref{sec_hdm}, or measuring time delays
of astrophysical neutrino bursts from $\gamma-$ray bursts and
supernovae relative to the speed of light.

\subsection{Big Bang Nucleosynthesis (BBN)}\label{sec_bbn}
For more detailed introductions to the following three topics
we refer the reader to standard text books~\cite{kt,mohapatra}.

\begin{figure}
\includegraphics[width=0.9\textwidth]{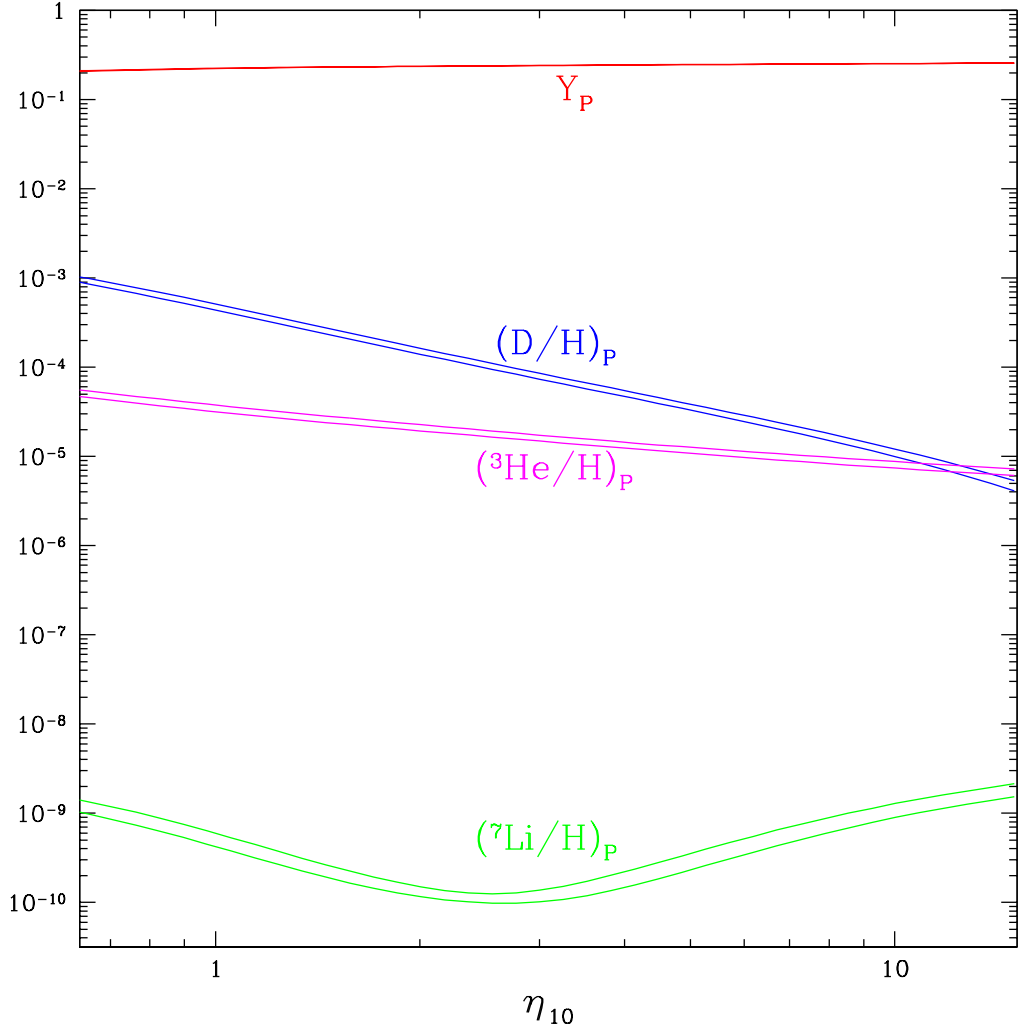}
\caption{From Ref.~\cite{steigman}. The SBBN-predicted primordial
abundances of D, $^3$He, and 
$^7$Li by number with respect to hydrogen, and the $^4$He mass fraction 
Y$_{\rm P}$, as a function of the nucleon (baryon) abundance parameter 
$\eta_{10}$.  The bands reflect the theoretical uncertainties 
($\pm 1\sigma$) in the BBN predictions.
\label{fig7}}
\end{figure}

The early universe consisted of a mixture of protons, neutrons,
electrons, positrons, photons and neutrinos. Their relative abundances
were determined
by thermodynamic equilibrium until the weak interactions "froze out"
once the temperature of the expanding universe dropped below
$T_f\sim1\,$MeV where their rates became smaller
than the expansion rate. For example, according to Eqs.~(\ref{cross})
and~(\ref{cross2}) the interaction rates of nucleons
$\overline{\nu}_e p\leftrightarrow ne^+$ and
$e^- p\leftrightarrow n\nu_e$ are
\begin{equation}
\Gamma\sim n\sigma\propto G_{\rm F}^2T^5\label{ewrates}
\end{equation}
at temperatures $100\,{\rm GeV}\ga T\ga1\,$MeV where the neutron-proton
mass difference $m_n-m_p=1.293\,$MeV and the electron mass are negligible and
the $e^\pm$ and electron neutrino densities $n\sim T^3$. This
becomes indeed comparable to the expansion rate
\begin{equation}
H\sim\rho^{1/2}/m_{\rm P}\sim g_*^{1/2}T^2/m_{\rm P}\,,\label{H}
\end{equation}
where $\rho$ is the total energy density and $g_*$ the number of
relativistic degrees of freedom, once $T$ approaches $T_f\simeq1\,$MeV.
The equilibrium neutron to proton ratio at that temperature is given by
thermodynamics as
\begin{equation}
  \frac{n_n}{n_p}=\exp\left[-(m_n-m_p)/T_f\right]
\end{equation}
At that time, the free neutrons were
quickly bound into helium which could not be broken up any more
by the cooling thermal radiation. The helium abundance was thus
determined by the freeze out of electroweak interactions. Since
equating Eq.~(\ref{ewrates}) with Eq.~(\ref{H}) yields
$T_f\propto g_*^{1/6}$ we also see that the helium abundance
should increase with $g_*$. Since the number $N_\nu$ of stable neutrino
species with mass below $\sim1\,$MeV contributes to $g_*$, this
number is constrained by the observed helium abundance. More
generally, in the absence of a significant asymmetry between
neutrinos and anti-neutrinos, elemental abundances depend only
on the effective number of relativistic neutrinos $N_\nu$ and
the baryon to photon ratio
\begin{equation}
  \eta_{10}\equiv10^{10}\frac{n_B}{n_\gamma}\,.
\end{equation}
Predictions for standard big bang nucleosynthesis (SBBN) with
$N_\nu=3$, the number of active neutrinos consistent with the
$Z$ boson width, are shown in Fig.~\ref{fig7}. A detailed comparison
of measured and predicted abundances shown in Fig.~\ref{fig7}
with $\eta_{10}$ and $N_\nu$ free parameters yields the following:
The universal density of baryons $\eta_{10}$ inferred from SBBN 
and the measured deuterium abundance, $\eta_{10}({\rm SBBN}) = 
6.10^{+0.67}_{-0.52}$, is in excellent agreement with the baryon 
density derived largely from CMB
data~\cite{spergel}, $\eta_{10}({\rm CMB}) = 6.14 \pm 0.25$.
However, there is a $\simeq2\sigma$ tension between the $^4$He
abundance predicted by SBBN with this concordance $\eta_{10}$
and the observed one. This tension can be mitigated if $N_\nu$ is
allowed to be smaller than the canonical $N_\nu=3$.
If {\it both} the baryon density $\eta_{10}$ and $N_\nu$, or
equivalently, the expansion rate, are allowed to be free parameters,
BBN (D, $^3$He, and $^4$He) and the CMB (WMAP) agree
at 95\% CL for $5.5 \leq \eta_{10} \leq 6.8$
($0.020 \leq \Omega_{\rm B}h^{2} \leq 0.025$ for the baryon density
in terms of the critical density) and
$1.65 \leq$~N$_{\nu} \leq 3.03$~\cite{steigman}.
Are these hints for new physics ?

\subsection{Neutrino Hot Dark Matter}\label{sec_hdm}
Finally, massive neutrinos in the eV range contribute to the density
of non-relativistic matter in today's universe,
\begin{equation}
  \Omega_\nu h^2=\frac{\sum m_\nu}{92.5\,{\rm eV}}\label{omeganu}
\end{equation}
in terms of the critical (closure) density for $m_\nu\gg10^{-4}\,$eV,
today's temperature. Eq.~(\ref{omeganu}) results from the fact that
neutrinos have been relativistic at decoupling at $T\sim1\,$MeV,
thus constituting {\it hot dark matter}, and their number density is
simply determined by the redshifted number density at freeze-out, in
analogy to Sect.~\ref{sec_bbn}.

Since neutrinos are
freely streaming on scales of many Mpc, the matter power spectrum
is reduced by a relative amount $\Delta P_m/P_m=-8\Omega_\nu/\Omega_m$, where
$\Omega_m$ is the total matter density. A combination of data on the
large scale structure and the CMB then leads to the limit~\cite{hannestad}
\begin{equation}
  \sum m_\nu\la 1.0\,{\rm eV}\,.\label{nucosmo}
\end{equation}
There was even a claim for a positive detection with~\cite{allen}
\begin{equation}
  \sum m_\nu\simeq 0.56\,{\rm eV}\,,
\end{equation}
but newest analyses suggest upper bounds even slightly below this~\cite{lya}.

It is intriguing that direct experimental bounds Eq.~(\ref{nudirect})
and cosmological bounds Eq.~(\ref{nucosmo}) have reached
comparable sensitivities. In addition, both a combination
of future CMB data from the Planck satellite with large scale
structure surveys~\cite{hannestad} and next generation laboratory
experiments such as KATRIN will probe the 0.1 eV regime.

Assuming three active neutrino oscillations with the parameters
discussed in Sects.~\ref{sec_sol} and~\ref{sec_atm} has an interesting
cosmological consequence: Flavor equilibrium is reached before
the BBN epoch and the asymmetry parameter $\xi_\nu=\mu_\nu/T$,
where $\mu_\nu$ is the common neutrino chemical potential, is
constrained by~\cite{dhpprs}
\begin{equation}
  \left|\xi_\nu\right|\la0.07\,.\label{xinu}
\end{equation}
As a consequence, neutrino degeneracy is unobservable in the
large scale structure and the CMB.

\subsection{Leptogenesis and Baryogenesis}
Neutrino masses may also play a key role in explaining the fact that
we live in a universe dominated by matter rather than anti-matter.
The heavy right-handed Majorana neutrinos involved in the seesaw
mechanism discussed in Sect.~\ref{sec_dirana} could have been produced in the
early Universe and their out-of-equilibrium
decays could give rise to a non-vanishing net lepton number $L$.
Non-perturbative quantum effects related to the non-abelian character
of the electroweak interactions can translate this into a net baryon
number $B$ while conserving $B-L$. The amount of baryon number
$n_B$ created in this scenario is related to the low-energy
leptonic $CP$-violation phase $\delta$~\cite{buchmueller}. Its
compatibility with the observed value for the
baryon per photon number $n_B/n_\gamma\simeq6\times10^{-10}$ implies
a lower bound $m_R\ga{\rm few}\,10^{10}\,$GeV. Via the see-saw relation
$m_\nu\simeq m_D^2/m_R$ for the light neutrino mass, this
corresponds to an optimal range $10^{-3}\,{\rm eV}\la m_\nu\la 0.1\,$eV,
in remarkable agreement with the observed atmospheric and solar neutrino
mass scales~\cite{buchmueller}. In general baryogenesis requires
violation of baryon number $B$, charge conjugation $C$, combined
charge and parity conjugation $CP$, and a departure from thermal
equilibrium, usually caused by the expansion of the Universe.
These conditions are known as the Sakharov conditions~\cite{sakharov}.
For more details see Refs.~\cite{mohapatra} and~\cite{kt}.

\section{Ultra-High Energy Cosmic Radiation}
In the final part we discuss some current theoretical issues around
ultra-high energy cosmic rays, $\gamma-$rays and neutrinos.
We will see how some of the topics discussed in the previous
two parts play an important role in this subject.

\subsection{Introduction}
High energy cosmic ray (CR) particles are shielded
by Earth's atmosphere and reveal their existence on the
ground only by indirect effects such as ionization and
showers of secondary charged particles covering areas up
to many km$^2$ for the highest energy particles. In fact,
in 1912 Victor Hess discovered CRs by measuring ionization from
a balloon~\cite{hess}, and in 1938 Pierre Auger proved the existence of
extensive air showers (EAS) caused by primary particles
with energies above $10^{15}\,$eV by simultaneously observing
the arrival of secondary particles in Geiger counters many meters
apart~\cite{auger_disc}.

\begin{figure}[ht]
\includegraphics[width=.9\textwidth,clip=true]{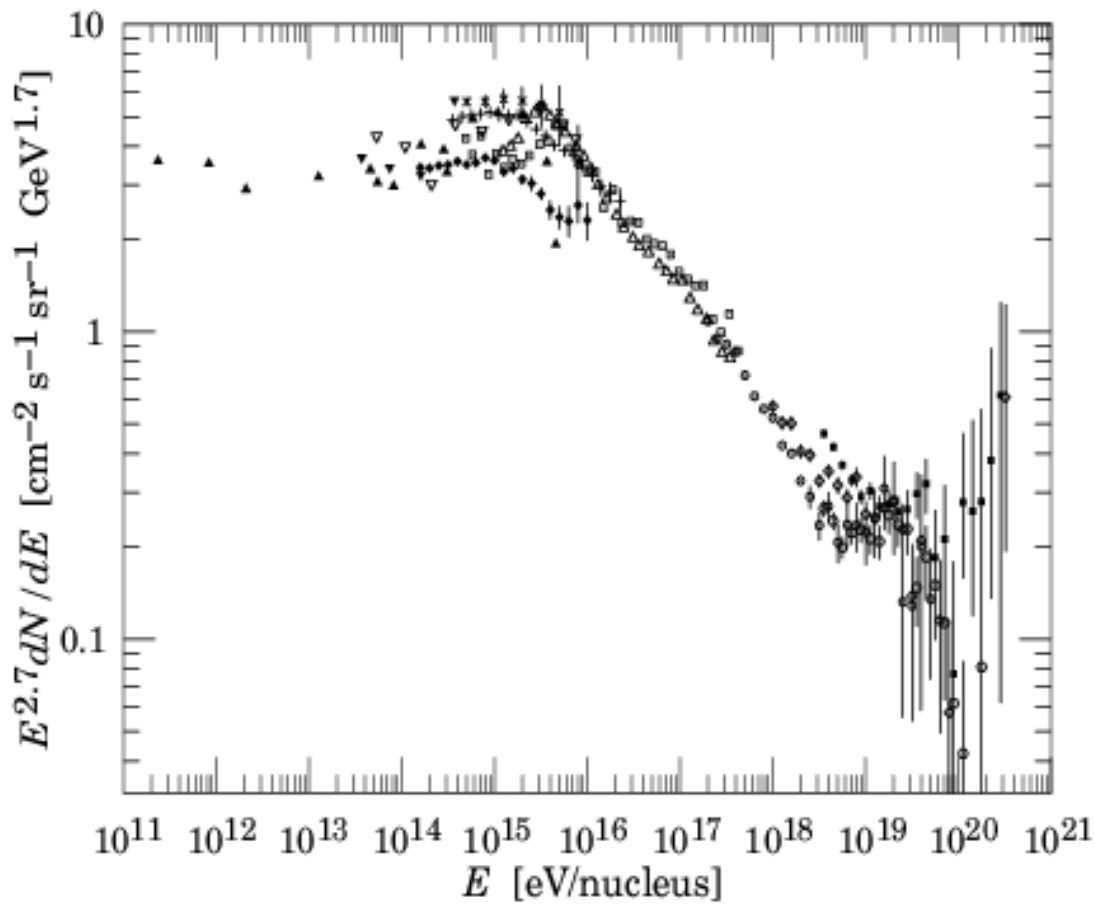}
\caption{From Ref.~\cite{rpp}. The cosmic ray all particle spectrum.
\label{fig8}}
\end{figure}

After almost 90 years of research, the origin of cosmic rays
is still an open question, with a degree of uncertainty
increasing with energy~\cite{crbook}: Only below 100 MeV
kinetic energy, where the solar wind shields protons coming
from outside the solar system, the sun must give rise to
the observed proton flux. Above that energy the CR spectrum
exhibits little structure and is approximated
by broken power laws $\propto E^{-\gamma}$:
At the energy $E\simeq4\times 10^{15}\,$eV
called the ``knee'', the flux of particles per area, time, solid angle,
and energy steepens from a power law index $\gamma\simeq2.7$
to one of index $\simeq3.0$. The bulk of the CRs up to at least
that energy is believed to originate within the Milky Way Galaxy,
typically by shock acceleration in supernova remnants.
The spectrum continues with a further steepening to $\gamma\simeq3.3$
at $E\simeq4\times 10^{17}\,$eV, sometimes called the ``second knee''.
There are experimental indications that the chemical composition
changes from light, mostly protons, at the knee to domination by
iron and even heavier nuclei at the second knee~\cite{kascade}.
This is in fact expected in any scenario where acceleration and
propagation is due to magnetic fields whose effects only depend
on rigidity, the ratio of charge to rest mass, $Z/A$. This is true
as long as energy losses and interaction effects, which in general depend
on $Z$ and $A$ separately, are small, as is the case in the Galaxy, in
contrast to extra-galactic cosmic ray propagation at ultra-high energy.
Above the so called ``ankle'' or ``dip'' at $E\simeq5\times10^{18}\,$eV, the
spectrum flattens again to a power law of index $\gamma\simeq2.8$.
This latter feature
is often interpreted as a cross over from a steeper Galactic
component, which above the ankle cannot be confined by the Galactic
magnetic field, to a harder component of extragalactic origin.
The dip at $E\simeq5\times10^{18}\,$eV could also be partially due
to pair production by extra-galactic protons, especially
if the extra-galactic component already starts to dominate below
the ankle, for example, around the second-knee~\cite{bgh}. This
latter possibility appears, however, less likely in light of
a rather heavy composition up to the ankle suggested by several
experiments~\cite{kascade}. In any case, an eventual cross over
to an extra-galactic component is also in line with experimental
indications for a chemical composition becoming again lighter above
the ankle, although a significant heavy component is not
excluded and the inferred chemical composition above
$\sim10^{18}\,$eV is sensitive to the model of air shower interactions
and consequently uncertain presently~\cite{watson}.
In the following we will restrict our discussion on ultra-high
energy cosmic rays (UHECRs) above the ankle where the spectrum seems
to continue up to several hundred EeV
(1 EeV$\equiv10^{18}\,$eV)~\cite{agasa,hires}, corresponding to about
50 Joules. The all-particle spectrum is shown in Fig.~\ref{fig8}.

We note that until the 1950s the energies achieved with experiments
at accelerators
were lagging behind observed CR energies which explains why many
elementary particles such as the positron, the muon, and the
pion were first discovered in CRs~\cite{battiston}. Today, where the
center of mass (CM) energies observed in collisions with atmospheric nuclei
reach up to a PeV, we have again a similar situation. In addition,
CR interactions in the atmosphere predominantly occur in the extreme
forward direction which allows to probe non-perturbative effects of
the strong interaction. This is complementary to collider experiments
where the detectors can only see interactions with significant
transverse momentum transfer.

Although statistically meaningful information about the UHECR energy
spectrum and arrival direction distribution has been accumulated, no
conclusive picture for the nature and distribution of the sources
emerges naturally from the data. There is on the one hand the approximate
isotropic arrival direction distribution~\cite{bm} which indicates that we are
observing a large number of weak or distant sources. On the other hand,
there are also indications which point more towards a small number of
local and therefore bright sources, especially at the highest energies:
First, the AGASA ground array claims statistically significant multi-plets of
events from the same directions within a few degrees~\cite{teshima1,bm},
although this is controversial~\cite{fw} and has not been seen so far
by the fluorescence experiment HiRes~\cite{finley}.
The spectrum of this clustered component is $\propto E^{-1.8}$ and thus
much harder than the total spectrum~\cite{teshima1}.
Second, nucleons above $\simeq70\,$EeV suffer heavy energy losses due to
photo-pion production on the cosmic microwave background
--- the Greisen-Zatsepin-Kuzmin (GZK) effect~\cite{gzk} already
mentioned in Sect.~\ref{sec_nuN} ---
which limits the distance to possible sources to less than
$\simeq100\,$Mpc~\cite{stecker}. For a uniform source distribution
this would predict a ``GZK cutoff'', a drop in the spectrum.
However, the existence of this ``cutoff'' is not established yet
from the observations~\cite{bergman} and may even depend on the
part of the sky one is looking at: The ``cutoff' could be mitigated
in the northern hemisphere where more nearby accelerators related
to the local supercluster can be expected. Apart from the SUGAR array
which was active from 1968 until 1979 in Australia, all UHECR detectors
completed up to the present were situated in the northern hemisphere.
Nevertheless the situation is unclear even there: Whereas a cut-off
seems consistent with the few events above $10^{20}\,$eV recorded
by the fluorescence detector HiRes~\cite{hires}, it is not compatible
with the 11 events above $10^{20}\,$eV measured by the AGASA ground
array~\cite{agasa}. It can be remarked, however, that analysis of
data based on a single fluorescence telescope, the so-called
monocular mode in which most of the HiRes data were obtained, is complicated
due to atmospheric conditions varying from event to event~\cite{cronin}.
The solution of this problem may have to await more analysis and,
in particular, the completion of the Pierre Auger project~\cite{auger}
which will combine the two complementary detection techniques
adopted by the aforementioned experiments and whose southern site
is currently in construction in Argentina.

This currently unclear experimental situation could easily be solved if it
would be possible to follow the UHECR trajectories backwards to their
sources. However, this may be complicated by the possible presence of
extragalactic magnetic fields, which would deflect the particles during
their travel. Furthermore, since the GZK-energy losses are of stochastic
nature, even a detailed knowledge of the extragalactic magnetic fields would
not necessarily allow to follow a UHECR trajectory backwards to its source
since the energy and therefor the Larmor radius of the particles
have changed in an
unknown way. Therefore it is not clear if charged particle astronomy with
UHECRs is possible in principle or not. And even if possible, it remains
unclear to which degree the angular 
resolution would be limited by magnetic deflection. This topic
will be discussed in Sect.~\ref{sec_egmf}.

\subsection{Severe Constraints on Scenarios producing more photons
than hadrons}\label{sec_egret}

\begin{figure}[ht]
\includegraphics[angle=270,width=.9\textwidth,clip=true]{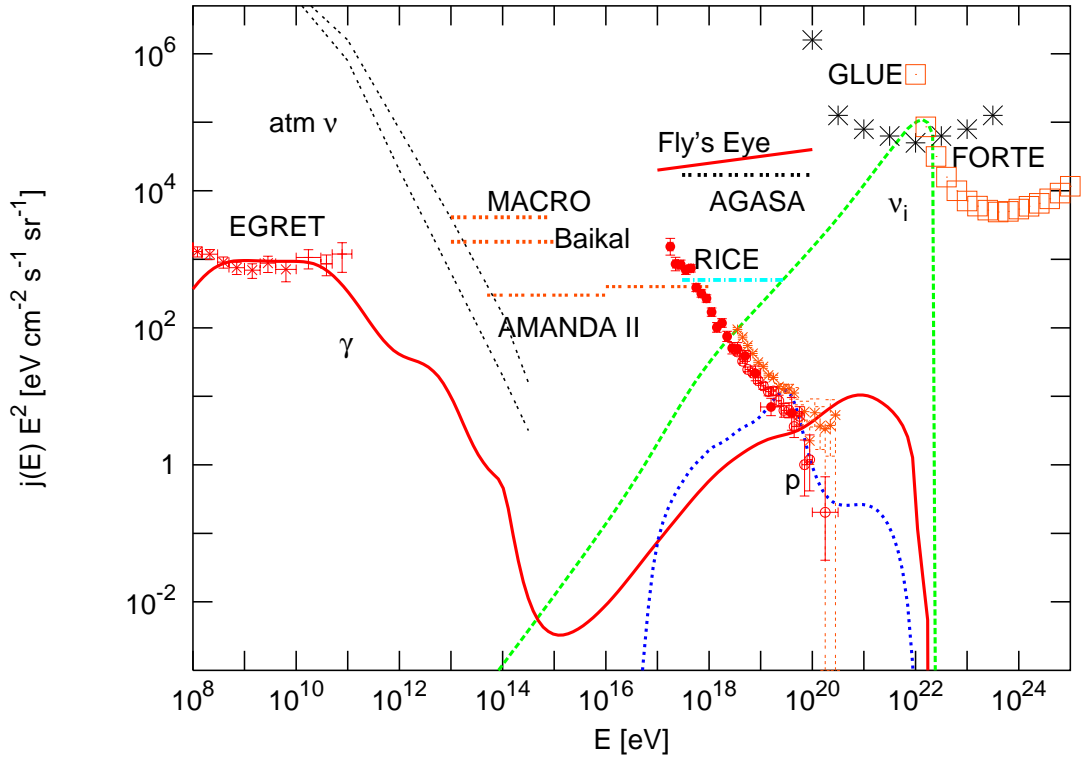}
\caption{From Ref.~\cite{ss}. Flux predictions for a Z-burst model
averaged over flavors and
characterized by a neutrino injection flux per comoving volume
$\propto E^{-1}$ up to $3\times10^{22}\,$eV and for redshifts between
0 and 3. The sources are assumed to be exclusive neutrino emitters. All
neutrino masses were assumed equal with $m_\nu=0.33$~eV and we again
assumed maximal mixing between all flavors. The data and upper limits
are as in Fig.~\ref{fig2}.
\label{fig9}}
\end{figure}

The physics and astrophysics of UHECRs are also intimately linked with
the emerging field of neutrino astronomy (for reviews see
Refs.~\cite{nu_review}) as well as with the already
established field of $\gamma-$ray astronomy (for reviews see, e.g.,
Ref.~\cite{gammarev}). Indeed, all
scenarios of UHECR origin, including the top-down models, are severely
constrained by neutrino and $\gamma-$ray observations and limits.
In turn, this linkage has important consequences for theoretical
predictions of fluxes of extragalactic neutrinos above about a TeV
whose detection is a major goal of next-generation
neutrino telescopes: If these neutrinos are
produced as secondaries of protons accelerated in astrophysical
sources and if these protons are not absorbed in the sources,
but rather contribute to the UHECR flux observed, then
the energy content in the neutrino flux can not be higher
than the one in UHECRs, leading to the so called Waxman-Bahcall
bound for transparent sources with soft acceleration
spectra~\cite{wb-bound,mpr}. This bound is shown in Fig.~\ref{fig2}.
If one of these assumptions does not apply, such as for acceleration
sources with injection spectra harder than $E^{-2}$ and/or opaque
to nucleons, or in the top-down scenarios where X particle decays
produce much fewer nucleons than $\gamma-$rays and neutrinos,
the Waxman-Bahcall bound does not apply, but the neutrino
flux is still constrained by the observed diffuse $\gamma-$ray
flux in the GeV range which is marked ``EGRET'' in Figs.~\ref{fig2}
and~\ref{fig9}. This bound whose implications will be discussed in
the following section is marked ``$\gamma-$ray bound'' in
Figs.~\ref{fig2} and~\ref{fig3}.

Electromagnetic (EM) energy injected above the threshold for
pair production on the CMB at
$\sim10^{15}/(1+z)\,$eV at redshift $z$ (to a lesser extent
also on the infrared/optical background, with lower threshold) leads
to an EM cascade, an interplay between pair production followed
by inverse Compton scattering of the produced electrons. This
cascade continues until the photons fall below the pair production
threshold at which point the universe becomes transparent for them.
In todays universe this happens within just a few Mpc for injection
up to the highest energies above $10^{20}\,$eV. All EM energy
injected above $\sim10^{15}\,$eV and at distances beyond a few
Mpc today is therefore recycled to lower energies where it gives
rise to a characteristic cascade spectrum $\propto E^{-2.1}$ down
to fractions of a GeV~\cite{bere}. The universe thus acts as a calorimeter
where the total EM energy injected above $\sim10^{15}/(1+z)\,$eV
is measured as a diffuse isotropic $\gamma-$ray flux in the GeV regime.
This diffuse flux is not very sensitive to the somewhat uncertain
infrared/optical background~\cite{ahacoppi}.
Any observed diffuse $\gamma-$ray background acts as an upper limit
on the total EM injection.
Since in any scenario involving pion production the EM energy fluence
is comparable to the neutrino energy fluence, the constraint
on EM energy injection also constrains allowed neutrino fluxes.

This diffuse extragalactic GeV $\gamma-$ray background can be
extracted from the total $\gamma-$ray flux measured by EGRET by
subtracting the Galactic contribution. Since publication of the
original EGRET limit in 1995~\cite{egret}, models for this high
latitude Galactic $\gamma-$ray foreground were improved
significantly. This allowed the authors of Ref.~\cite{egret_new}
to reanalyze limits on the diffuse extragalactic background
in the region 30 MeV-10 GeV and to lower it by a factor 1.5-1.8
in the region around 1 GeV. There are even lower
estimates of the extragalactic diffuse $\gamma-$ray flux~\cite{kwl}.
In this article, however, we will use the more conservative 
limits from Ref.\cite{egret_new}.

The energy in the extra-galactic $\gamma-$ray background estimated
in Ref.~\cite{egret_new} is slightly more than one hundred times the
energy in UHECR above the GZK cutoff. The range of such trans-GZK
cosmic rays is about $\simeq30\,$Mpc, roughly one hundredth the
Hubble radius, and only sources within that GZK range contribute
to the trans-GZK cosmic rays. Therefore, any mechanism involving
sources distributed roughly uniformly on scales of the GZK energy
loss length $\simeq30\,$Mpc and producing a comparable amount of energy
in trans-GZK cosmic rays and photons above the pair production threshold
can potentially explain this energy flux ratio. The details depend
on the exact redshift dependence of source activity and other
parameters and in general have to be verified by numerically solving
the relevant transport equations, see, e.g., Ref.~\cite{ss}. Such
mechanisms include shock acceleration in powerful objects such as
active galactic nuclei~\cite{ta}.

On the other hand, any mechanism producing considerably {\it more}
energy in the EM channel above the pair production threshold than
in trans-GZK cosmic rays tend to predict a ratio of the diffuse
GeV $\gamma-$ray flux to the trans-GZK cosmic ray flux too high
to explain both fluxes at the same time. As a consequence, if
normalized at or below the observational GeV $\gamma-$ray background, such
scenarios tend to explain at most a fraction of the observed
trans-GZK cosmic ray flux. Such scenarios include particle physics
mechanisms involving pion production by quark fragmentation, e.g.
extra-galactic top-down mechanisms where UHECRs are produced by
fragmenting quarks resulting from decay of superheavy relics~\cite{bs-rev}.
Most of these quarks would fragment into pions rather than nucleons
such that more $\gamma-$rays (and neutrinos) than cosmic rays
are produced. Overproduction of GeV $\gamma-$rays can be avoided
by assuming the sources in an extended Galactic halo with a high $\ga10^3$
overdensity compared to the average cosmological source density, which
would also avoid the GZK cutoff~\cite{bkv}.
These scenarios, however, start to be constrained by the anisotropy
they predict because of the asymmetric position of the Sun
in the Galactic halo for which there are no indications in present
data~\cite{ks2003}. Scenarios based on quark fragmentation also become
problematic in view of a possible heavy
nucleus component and of upper limits on the photon fraction of
the UHECR flux~\cite{watson}.

As a specific example for scenarios involving quark fragmentation,
we consider here the case of decaying
Z-bosons. In this ``Z-burst mechanism'' Z-bosons are produced by
UHE neutrinos interacting with the relic neutrino
background~\cite{zburst1}. If the relic neutrinos
have a mass $m_\nu$, Z-bosons can be resonantly produced by UHE
neutrinos of energy
$E_\nu\simeq M_Z^2/(2m_\nu)\simeq4.2\times10^{21}\,{\rm eV}\,({\rm eV}/m_\nu)$.
The required neutrino
beams could be produced as secondaries of protons accelerated
in high-redshift sources. The fluxes predicted in these scenarios
have recently been discussed in detail, for example, in Refs.~\cite{fkr,ss}.
In Fig.~\ref{fig9} we show an optimistic example taken from Ref.~\cite{ss}.
It is assumed that the relic neutrino background has no significant
local overdensity. Furthermore, the sources
are assumed to not emit any $\gamma-$rays, otherwise the Z-burst
model with acceleration sources over-produces the diffuse GeV $\gamma-$ray
background~\cite{kkss}. We note that no known
astrophysical accelerator exists that meets the requirements
of the Z-burst model~\cite{kkss,gtt2003}.

However, a combination of new constraints discussed in the
previous sections allows to rule out that the Z-burst mechanism
explains a dominant fraction of the observed UHECR flux, even for
pure neutrino emitting sources: As discussed in Sect.~\ref{sec_hdm},
a combination of cosmological data including the WMAP experiment limit
the sum of the masses of active neutrinos to $\la1\,$eV~\cite{hannestad}.
In Sects.~\ref{sec_sol},~\ref{sec_atm} we have seen that solar and
atmospheric neutrino oscillations indicate that individual
neutrino masses are nearly degenerate on this scale~\cite{maltoni}, and thus
the neutrino mass per flavor must satisfy $m_\nu\la0.33\,$eV.
However, for such masses phase space constraints limit the possible
over-density of neutrinos in our Local Group of galaxies to
$\la10$ on a length scale of $\sim1\,$Mpc~\cite{sm}. Since
this is considerably smaller than the relevant UHECR loss lengths,
neutrino clustering will not significantly reduce the necessary UHE neutrino
flux compared to the case of no clustering.
For the maximal possible value of the neutrino
mass $m_\nu \simeq 0.33\,$eV, the neutrino flux required for the Z-burst
mechanism to explain the UHECR flux is only in marginal conflict with the
FORTE upper limit~\cite{forte}, and factor 2 higher than the new GLUE
limit~\cite{glue}, as shown in Fig.~\ref{fig9}. For all other
cases the conflict with both the GLUE and FORTE limits is
considerably more severe. 
Also note that this argument does not depend on the shape of the low energy
tail of the primary neutrino spectrum which could thus be even
mono-energetic, as could occur in exclusive tree level decays
of superheavy particles into neutrinos~\cite{gk}. However, in addition
this possibility has been ruled out by overproduction of GeV
$\gamma-$rays due to loop effects in these particle decays~\cite{bko}.

The possibility that the
observed UHECR flux is explained by the Z burst scenario involving
normal astrophysical sources which produce both neutrinos and photons
by pion production is already ruled out by the former EGRET limit:
In this case the GeV $\gamma-$ray flux level would have roughly
the height of the peak of the neutrino flux multiplied with the
squared energy in Fig.~\ref{fig9}, thus a factor $\sim100$ higher
than the EGRET level.

Any further reduction in the estimated contribution of the true diffuse
extra-galactic $\gamma-$ray background to the observed flux, therefore,
leads to more severe constraints on the total EM injection.
For example, future $\gamma-$ray detectors such as GLAST~\cite{glast}
will test whether the diffuse extragalactic GeV $\gamma-$ray background
is truly diffuse or partly consists of discrete sources that could
not be resolved by EGRET. Astrophysical discrete contributions such
as from intergalactic shocks are in fact expected~\cite{astrocontr}.
This could further improve the cascade limit
to the point where even acceleration scenarios may become seriously
constrained.

\subsection{New Primary Particles}

A possible way around the problem of missing counterparts
within acceleration scenarios is to propose primary
particles whose range is not limited by interactions with
the CMB. Within the Standard Model the only candidate is the neutrino,
whereas in extensions of the Standard Model one could think of
new neutrals such as axions or stable supersymmetric elementary
particles. Such options are mostly ruled out by the tension
between the necessity of a small EM coupling to avoid the GZK cutoff and
a large hadronic coupling to ensure normal air showers~\cite{ggs}.
Also suggested have been
new neutral hadronic bound states of light gluinos with
quarks and gluons, so-called R-hadrons that are heavier than
nucleons, and therefore have a higher GZK threshold~\cite{cfk},
as can be seen from Eq.~(\ref{gzk}).
Since this too seems to be disfavored by accelerator
constraints~\cite{gluino} we will here focus on neutrinos.

In both the neutrino and new neutral stable particle scenario
the particle propagating over extragalactic distances would have
to be produced as a secondary in interactions of a primary proton
that is accelerated in a powerful active galactic nucleus
which can, in contrast to the case of EAS induced by nucleons,
nuclei, or $\gamma-$rays,
be located at high redshift. Consequently, these scenarios predict
a correlation between primary arrival directions
and high redshift sources. In fact, possible evidence
for a correlation of UHECR arrival directions with compact radio
quasars and BL-Lac objects, some of them possibly too far away to
be consistent with the GZK effect, was recently reported~\cite{bllac}.
The main challenge in these correlation studies is the choice
of physically meaningful source selection criteria and the avoidance
of a posteriori statistical effects. However, a moderate increase
in the observed number of events will most likely confirm or
rule out the correlation hypothesis. Note, however, that
these scenarios require the primary proton to be accelerated
up to at least $10^{21}\,$eV, demanding a very powerful
astrophysical accelerator.

\subsection{New Neutrino Interactions}

Neutrino primaries have the advantage of being well established
particles. However, within the Standard Model their interaction cross
section with nucleons shown in Fig.~\ref{fig1} falls short by about
five orders of magnitude to produce air showers starting high in
the atmosphere as observed. Electroweak instantons could change
this but this possibility is speculative~\cite{ew_instanton}.
The neutrino-nucleon cross section, $\sigma_{\nu N}$, however, can
be enhanced by new physics beyond the electroweak scale in the
CM frame, or above about a PeV in the nucleon rest frame.
Note that the CM energy reached by an UHECR nucleon of energy
$E$ interacting with an atmospheric nucleon at rest is
$\sqrt s\simeq0.4(E/10^{20}\,{\rm eV})\,$PeV.
Neutrino induced air showers may therefore rather directly
probe new physics beyond the electroweak scale.

One possibility consists of a large increase
in the number of degrees of freedom above the electroweak 
scale~\cite{kovesi-domokos}. A specific instance
of this idea appears in theories with $n$ additional large
compact dimensions and a quantum gravity scale $M_{4+n}\sim\,$TeV
that has recently received much attention in the literature~\cite{tev-qg}
because it provides an alternative solution to the hierarchy problem
in grand unifications of gauge interactions without a need
of supersymmetry. The idea is to dimensionally reduce the
$n+4$ dimensional gravitational action
\begin{equation}
  S_g=-\frac{M_{4+n}^{2+n}}{16\pi}\int d^{4+n}x\sqrt{-g}R\,,\label{eh}
\end{equation}
with $g$ the determinant of the metric and $R$ the Ricci scalar,
to four dimensions by integrating out the $n$ compact dimensions.
This yields the relation
\begin{equation}
  M_{\rm Pl}^2=V_n M_{4+n}^{2+n}\,,
\end{equation}
where $M_{\rm Pl}$ and $V_n$ are the four-dimensional Planck mass
and the volume of the $n$ extra dimensions, respectively. The
weakness of gravity can now be understood as a consequence of the
fact that it is the only force that propagates into the extra
dimensions: Their large volume dilutes gravitational interactions
between the Standard Model particles which are confined to a
3-brane representing our world. For compact extra dimensions,
gravity would only be modified at scales below
\begin{equation}
  r_n\simeq M^{-1}_{4+n}\left(\frac{M_{\rm Pl}}{M_{4+n}}\right)^{2/n}
  \simeq2\times10^{-17}\left(\frac{{\rm TeV}}{M_{4+n}}\right)
  \left(\frac{M_{\rm Pl}}{M_{4+n}}\right)^{2/n}\,{\rm cm}
  \,,\label{rextra}
\end{equation}
which is $r_n\la1\,$mm for $n\geq2$, $M_{4+n}\ga\,$TeV, and thus
consistent with gravity tests at small distances. In contrast,
non-gravitational interactions are
confined to the 3-brane and thus the Standard Model is not modified.

The neutrino-nucleon cross section in these frameworks is obtained
by substituting the new fundamental cross section $\sigma_i(xs,Q)$
for the electroweak cross section at the parton level in Eq.~(\ref{nuX}),
\begin{equation}
  \frac{d\sigma^{\nu X}}{dxdy}=
  \sum_i f_i(x,Q)\sigma_i(xs,Q)\,.\label{nuXnew}
\end{equation}
One of the largest contributions to the neutrino-nucleon
cross section turns out to be the production on our 3-brane
of microscopic black holes which are solutions of $4+n$-dimensional
gravity described by Eq.~(\ref{eh}).
These cross sections scale as $(s/M_{4+n}^2)^{1/(n+1)}$
for $s\ga M_{4+n}^2$. Their UV-divergence is due to the non-renormalizable,
classical character of the gravitational interaction Eq.~(\ref{eh}) in
the sense of Sect.~\ref{sec_renorm}. The production of
compact branes, completely wrapped around the extra dimensions,
may provide even larger contributions~\cite{aco}. The resulting
total cross sections can be larger than in the Standard Model by up to
a factor $\sim100$ if $M_{4+n}\sim\,$TeV~\cite{fs}. However, extra
dimensions with a flat geometry are severely constrained by
astrophysics: Core collapse of massive stars would lead to
production of gravitational excitations in the large compact
extra dimensions, mostly by nucleon-nucleon bremsstrahlung. These so called
Kaluza-Klein gravitons of mass $m_g$ are then gravitationally
trapped around the newly born neutron star during their livetime
$\tau\sim M_{\rm Pl}^2/m_g^3\sim10^{13}(10\,{\rm MeV}/m_g)^3\,$yr.
Their subsequent decay into two $\gamma-$rays
would make neutron stars shine in $\gamma-$rays. The non-observation
of such emission leads to lower bounds on $M_{4+n}$ which
decrease with increasing $n$, starting with $M_5\ga10^5\,$TeV and going
down to $M_9\ga1\,$TeV and still lower values for larger
$n$~\cite{hannestad1,casse}, for flat compact extra dimensions.
This implies that significant contributions to the neutrino-nucleon
cross section in these extra dimension scenarios require
either $n\geq5$ extra dimensions or a warped geometry.

Whereas the sub-hadronic scale cross sections obtained in
some extra dimension scenarios are still too small to be consistent
with observed air showers and thus to explain
the observed UHECR events~\cite{kp}, they can still have
important phenomenological consequences. This is because
UHECR data can be used to put constraints on cross sections
satisfying $\sigma_{\nu N}(E\ga10^{19}\,{\rm eV})
\la10^{-27}\,{\rm cm}^2$. Particles with such cross
sections would give rise to horizontal air showers which have
not yet been observed. Resulting upper limits on their fluxes
assuming the Standard Model cross section Eq.~(\ref{cross}) are
shown in Fig.~\ref{fig2}. Comparison with the ``cosmogenic'' neutrino flux
produced by UHECRs interacting with the CMB then results in upper
limits on the cross section which are about a factor 1000 larger
than Eq.~(\ref{cross}) in the energy range between $\simeq10^{17}\,$eV
and $\simeq10^{19}\,$eV~\cite{mr,tol,afgs}. The projected sensitivity of
future experiments shown in Fig.~\ref{fig3} indicate that these limits
could be lowered down to the Standard Model cross section~\cite{afgs}. In case
of a detection of penetrating events the degeneracy of the cross
section with the unknown neutrino flux could be broken by comparing the rates
of horizontal air showers with the ones of Earth skimming events~\cite{kw}.
This would allow to ``measure'' the neutrino-nucleon cross section
at energies unreachable by any forseeable terrestrial accelerator !

\subsection{Violation of Lorentz Invariance}\label{sec_vli}
The most elegant solution to the problem of apparently missing nearby
sources of UHECRs and for their putative correlation with high redshift
sources would be to speculate that the GZK effect does not exist theoretically.
A number of authors pointed out~\cite{vli_others,cg} that this may be possible
by allowing violation of Lorentz invariance
(VLI) by a tiny amount that is consistent with all current
experiments. At a purely theoretical level, several quantum gravity models
including some based on string theories do in fact predict non-trivial
modifications of space-time symmetries that also imply VLI at extremely
short distances (or equivalently at extremely high energies); see e.g., 
Ref.~\cite{amelino-piran} and references therein. These theories 
are, however, not yet in forms definite enough to allow precise
quantitative predictions of the exact form of the possible VLI. 
Current formulations of the effects of a possible VLI on high energy
particle interactions relevant in the context of UHECR, therefore, adopt a 
phenomenological approach in which the form of the possible
VLI is parametrized in various
ways. VLI generally implies the existence of a universal preferred frame
which is usually identified with the frame that is
comoving with the expansion of the Universe, in which the CMB
is isotropic. 

A direct way of introducing VLI is through a modification
of the standard {\it dispersion relation},
$E^2-p^2=m^2$, between energy $E$ and momentum $p=|\vec{p}|$ of particles, $m$
being the invariant mass of the particle. Currently there is no 
unique way of parameterizing the possible modification of this relation in
a Lorentz non-invariant theory. We discuss here a parameterization
of the modified dispersion relation which covers most of the qualitative
cases discussed in the literature and, for certain parameter values,
allows to completely evade the GZK limit,
\begin{equation}
  E^2-p^2-m^2\simeq-2dE^2-\xi\frac{E^3}{M_{\rm Pl}}-
  \zeta\frac{E^4}{M^2_{\rm Pl}}\,.\label{vli1}
\end{equation}
Here, the Planck mass $M_{\rm Pl}$ characterizes non-renormalizable effects
with dimensionless coefficients $\xi$ and $\zeta$, and the dimensionless
constant $d$ exemplifies VLI effects due to renormalizable terms
in the Lagrangian, see the discussion in Sect.~\ref{sec_renorm}. The standard
Lorentz invariant dispersion relation is recovered in the limit
$\xi,\zeta,d\to0$.

The constants $d\neq0$ can break Lorentz invariance spontaneously when
certain Lorentz tensors $c_{\mu\nu}$ have couplings to fermions of the form
$d_{\mu\nu}\bar\psi\gamma^\mu\partial^\nu\psi$, and acquire
vacuum expectation values of the form
$\langle d_{\mu\nu}\rangle=d\delta_\mu^0\delta_\nu^0$. If rotational
invariance and gauge symmetry are preserved, such renormalizable
Lorentz invariance breaking terms in the Lagrangian, whose Lorentz
invariant part is given by Eq.~(\ref{L_matter}),
are characterized by a single time-like vector $u^\mu$, with
$u^\mu u_\mu=-1$, which defines
a preferred reference frame~\cite{ck}.
The dimensionless terms can be interpreted as
a change of the maximal particle velocity~\cite{cg}
$v_{\rm max}=\partial E/\partial p|_{E,p\gg m}\simeq1-d$. At a fixed
energy $E$ one has the correspondence $d\to(\xi/2)(E/M_{\rm Pl})+
(\zeta/2)(E/M_{\rm Pl})^2$, as can be seen from Eq.~(\ref{vli1}).

Within effective field theory, effects of first order in $M^{-1}_{\rm Pl}$,
$\xi\neq0$, arise from the most general terms of the form
\begin{equation}
  \frac{\kappa}{2M_{\rm Pl}}u^\mu F_{\mu\nu}(u\cdot\partial)u_\lambda
  \tilde{F}^{\lambda\nu}+
  \frac{1}{2M_{\rm Pl}}u^\mu\bar\psi\gamma_\mu(\lambda_1+\lambda_2\gamma_5)
  (u\cdot\partial)^2\psi\,,\label{vlifirst}
\end{equation}
where $\tilde{F}$ denotes the dual of the field strength $F$. For
photons and electrons this leads to $\xi=\pm\kappa$ and\
$\xi=\lambda_1\pm\lambda_2$, respectively, in Eq.~(\ref{vli1}),
where $\pm$ refers to helicity which remains conserved in the
presence of the terms Eq.~(\ref{vlifirst})~\cite{pospelov}. These
terms also violate $CPT$. Effects of first order in $M^{-1}_{\rm Pl}$,
$\xi\neq0$, are possible, for example, in non-critical Liouville
string theory due to recoiling D-branes~\cite{emn}. Finally, in
critical string theory,
effects second order in $M^{-1}_{\rm Pl}$, $\zeta\neq0$, can be induced
due to quantum gravity effects.

Interestingly, it has been pointed out recently that in the
supersymmetric Standard
Model VLI terms must be non-renormalizable and do not lead to
modifications of any dispersion relations~\cite{nibbelink}.

Now, consider the GZK photo-pion production process in
which a nucleon of energy $E$, momentum $p$ and mass $m_N$ collides
head-on with a CMB photon of energy $\epsilon$ producing a pion and a
recoiling nucleon. The threshold initial momentum of the nucleon for this
process according to standard Lorentz invariant kinematics is 
\begin{equation}
p_{{\rm th},0}=(m_{\pi}^{2}+2m_{\pi}m_N)/4\epsilon\,,\label{gzk-th} 
\end{equation}
where $m_\pi$ and $m_N$ are the pion and nucleon masses, respectively.
Assuming exact energy-momentum conservation but using the modified
dispersion relation given above, in the
ultra-relativistic regime $m\ll p\ll M$, and neglecting sub-leading terms,
the new nucleon threshold momentum $p_{\rm th}$ under the modified
dispersion relation Eq.~(\ref{vli1}) for $d=0$ satisfies~\cite{aloisio} 
\begin{equation} 
-\beta x^4-\alpha x^3 + x - 1 = 0\,,
\label{vli2} 
\end{equation}
where $x=p_{\rm th}/p_{{\rm th},0}$, and 
\begin{eqnarray} 
\alpha&=&\frac{2\xi p_{{\rm th},0}^{3}}{(m_{\pi}^{2}+2m_\pi m_N)M_{\rm Pl}}
 \frac{m_\pi m_N}{(m_\pi + m_N)^2}\,,\label{vli3}\\
\beta&=&\frac{3\zeta p_{{\rm th},0}^{4}}{2(m_{\pi}^{2}+2m_\pi m_N)M^2_{\rm Pl}}
 \frac{m_\pi m_N}{(m_\pi + m_N)^2}\,.\nonumber
\end{eqnarray} 
One can show that the same modified dispersion
relation Eq.~(\ref{vli1}) leads to the same condition Eq.~(\ref{vli2})
for absorption of high energy gamma rays through $e^+e^-$ pair production
on the infrared, microwave or radio backgrounds, if one substitutes
$p_{{\rm th},0}=m^2_e/\epsilon$,
$\alpha=\xi p_{{\rm th},0}^3/(8m^2_e M_{\rm Pl})$,
$\beta=3\zeta p_{{\rm th},0}^4/(16m^2_e M^2_{\rm Pl})$, where $m_e$
is the electron mass.

If $\xi,\zeta\simeq1$, there is no real positive solution of
Eq.~(\ref{vli2}), implying that the GZK process does not take place and
consequently the GZK cutoff effect disappears completely. Thus UHE
nucleons and/or photons will be able to reach Earth from any distance. 
On the other hand, if future UHECR data confirm the presence of a GZK
cutoff at some energy then that would imply upper limits on the
couplings $\xi$ and $\zeta$, thus probing specific
Lorentz non-invariant theories. If $p_{\rm th}\simeq p_{{\rm th},0}$,
one could conclude from Eq.~(\ref{vli2}) that $\alpha,\beta\la1$,
which translates into $|\xi|\la10^{-13}$ for the first order effects,
and $|\zeta|\la10^{-6}$ for the second order
effects, $\xi=0$~\cite{aloisio}. These values correspond to values
$|d|\la10^{-23}$ for the paremeters of renormalizable VLI.
Confirmation of a cut-off for
TeV photons with next-generation $\gamma-$ray observatories
would lead to somewhat weaker constraints~\cite{tev}.

More generally, modification of reaction kinematics or new
reaction channels are expected whenever the terms on the right hand
side of Eq.~(\ref{vli1}) become comparable to $m^2$, in rough numbers,
\begin{eqnarray}
  d&\ga&\frac{m^2}{2E^2}\simeq5\times10^{-23}
  \left(\frac{m}{{\rm GeV}}\right)^2
  \left(\frac{E}{10^{20}{\rm eV}}\right)^{-2}\,,\nonumber\\
  \xi&\ga&\frac{M_{\rm Pl}m^2}{E^3}\simeq10^{-14}
  \left(\frac{m}{{\rm GeV}}\right)^2
  \left(\frac{E}{10^{20}{\rm eV}}\right)^{-3}\,,\label{vlirough}\\
  \zeta&\ga&\frac{M_{\rm Pl}^2m^2}{E^4}\simeq10^{-6}
  \left(\frac{m}{{\rm GeV}}\right)^2
  \left(\frac{E}{10^{20}{\rm eV}}\right)^{-4}\,.\nonumber
\end{eqnarray}
Note that by far the smallest parameter values would be probed
by particles with the smallest mass, specifically the neutrino,
$m\la\,$eV at the highest energies. This makes the prospects of
future detections of cosmogenic neutrinos, see, e.g., Fig.~\ref{fig2},
very exciting also for VLI constraints.

In addition, the non-renormalizable terms in the dispersion relation
Eq.~(\ref{vli1}) imply a change in the group velocity which for the
first-order term leads to time delays over distances $r$ given by
\begin{equation}
  \Delta t\simeq\xi r\frac{E}{M_{\rm Pl}}\simeq\xi
  \left(\frac{r}{100\,{\rm Mpc}}\right)\left(\frac{E}{{\rm TeV}}\right)
  \,{\rm sec}\,.\label{delay} 
\end{equation}
For $|\xi|\sim1$ such time delays could be measurable, for example,
by fitting the arrival times of $\gamma-$rays arriving from
$\gamma-$ray bursts to the predicted energy dependence.

We mention that if VLI is due to modification of the
space-time structure expected in some theories of quantum gravity,
for example, then the strict energy-momentum conservation assumed in the above
discussion, which requires space-time translation invariance, 
is not guaranteed in general, and then the calculation of the modified
particle interaction thresholds becomes highly non-trivial and
non-obvious. Also, it is possible that a Lorentz non-invariant
theory while giving a  modified dispersion relation also imposes
additional kinematic structures such as a modified law of addition
of momenta. Indeed, Ref.~\cite{amelino-piran} gives an example of
a so-called $\kappa$-Minkowski non-commutative space-time in which the modified
dispersion relation has the same form as in Eq.~(\ref{vli1}) but there is also
a modified momentum addition rule which compensates for the effect of the
modified dispersion relation on the particle interaction thresholds
discussed above leaving the threshold momentum unaffected and consequently  
the GZK problem unsolved. In scenarios where the relativity of
inertial frames is preserved by a non-linear representation of the
Poincar\'e group, thresholds are in general significantly modified
only if the effective mass scale $M_{\rm Pl}/\xi$ is of the order
of the unmodified threshold energy in the laboratory frame~\cite{ms}.

There are several other fascinating effects of allowing a small VLI,
some of which are relevant for the question of origin and propagation
of UHECR. For example, any movement relative to the preferred frame
defined by $u^\mu$ in Eq.~(\ref{vlifirst}) gives rise to spatial
anisotropy. Clock comparison and spin precession experiments then
lead to limits on the dimensionless parameters in Eq.~(\ref{vlifirst})
between ${\cal O}(1)$ and ${\cal O}(10^{-8})$, depending on the
particle~\cite{pospelov}. Similar limits result from astrophysical
arguments: The observation of polarized MeV synchrotron radiation
from electrons in the Crab nebula implies the absence of vacuum
\v{C}erenkov radiation $e\to e\gamma$ for electrons up to energies
$E\sim1.5\,$PeV~\cite{jlms,jlm}. This process can become possible
if the electron speed becomes larger than the speed of light at
high energies and leads to limits on VLI of size comparable to the
before mentioned laboratory constraints~\cite{jlms,jlm}.
These constraints basically rule out effects of order $E/M_{\rm Pl}$
which might be a challenge for certain quantum gravity scenarios~\cite{pr}.
Note that these current constraints on VLI parameters still allow
strong modification of GZK kinematics by VLI parameters of the order
given in Eq.~(\ref{vlirough}).

\subsection{Cosmic Magnetic Fields and Their Influence on Ultra-High
Energy Cosmic Ray Propagation}\label{sec_egmf}

Cosmic magnetic fields are inextricably linked with cosmic rays
in several respects. First, they play a central role in Fermi
shock acceleration. Second, large scale extra-galactic magnetic
fields (EGMF) can cause significant deflection of charged cosmic
rays during propagation and thus obviously complicate the relation
between observed UHECR distributions and their sources.

Magnetic fields are omnipresent in the Universe, but their
true origin is still unclear~\cite{bt_review}. Magnetic fields
in galaxies are observed with typical strengths of a few
micro Gauss, but there are also some indications for fields correlated
with larger structures such as galaxy clusters~\cite{bo_review}.
Magnetic fields as strong as
$\simeq 1 \mu G$ in sheets and filaments of the large scale galaxy
distribution, such as in our Local Supercluster, are compatible with
existing upper limits on Faraday rotation~\cite{bo_review,ryu,blasi}.
It is also possible that fossil cocoons of former radio galaxies,
so called radio ghosts, contribute significantly to the isotropization
of UHECR arrival directions~\cite{mte}.

To get an impression of typical deflection angles one can characterize the
EGMF by its r.m.s. strength $B$ and a coherence length $l_c$.
If we neglect energy loss processes for the moment, then
the r.m.s. deflection angle over a distance $r\ga l_c$ in such a field
is $\theta(E,r)\simeq(2rl_c/9)^{1/2}/r_L$~\cite{wm}, where the Larmor
radius of a particle of charge $Ze$ and energy $E$ is
$r_L\simeq E/(ZeB)$. In numbers this reads
\begin{equation}
  \theta(E,r)\simeq0.8^\circ\,
  Z\left(\frac{E}{10^{20}\,{\rm eV}}\right)^{-1}
  \left(\frac{r}{10\,{\rm Mpc}}\right)^{1/2}
  \left(\frac{l_c}{1\,{\rm Mpc}}\right)^{1/2}
  \left(\frac{B}{10^{-9}\,{\rm G}}\right)\,,\label{deflec}
\end{equation}
for $r\ga l_c$. This expression makes it immediately obvious
that fields of fractions of micro Gauss lead to strong deflection
even at the highest energies.
This goes along with a time delay $\tau(E,r)\simeq r\theta(E,d)^2/4
\simeq1.5\times10^3\,Z^2(E/10^{20}\,{\rm eV})^{-2}
(r/10\,{\rm Mpc})^{2}(l_c/{\rm Mpc})(B/10^{-9}\,{\rm G})^2\,$yr
which can be millions of years. A source visible in UHECRs today
could therefore be optically invisible since many models involving,
for example, active galaxies as UHECR accelerators, predict
variability on shorter time scales.

Quite a few simulations of the effect of extragalactic magnetic fields
(EGMF) on UHECRs exist in the literature, but usually idealizing
assumptions concerning properties and distributions of sources
or EGMF or both are made: In Refs.~\cite{slb,ils,lsb,sse,is} sources
and EGMF follow a pancake profile mimicking the local supergalactic
plane. In other studies EGMF have been approximated
in a number of fashions: as negligible~\cite{sommers,bdm},
as stochastic with uniform statistical properties~\cite{bo,ynts,ab},
or as organized in spatial cells with a given coherence length and a strength
depending as a power law on the local density~\cite{tanco}.
Only recently attempts have been made to simulate UHECR propagation
in a realistically structured universe~\cite{sme,dolag}. For
now, these simulations are limited to nucleons.

In Ref.~\cite{sme} the magnetized extragalactic environment used
for UHECR propagation is produced by a simulation of the large scale
structure of the Universe. The simulation was carried out
within a computational box of $50\,h^{-1}\,$Mpc length on a side, 
with normalized Hubble constant 
$h\equiv H_0/(100$ km s$^{-1}$ Mpc$^{-1})$ = 0.67, and using
a comoving grid of 512$^3$ zones and 256$^3$ dark matter
particles. The EGMF was initialized to zero at simulation start 
and subsequently its seeds were
generated at cosmic shocks through the Biermann battery
mechanism~\cite{kcor97}. Since cosmic shocks form
primarily around collapsing structures including filaments, the above
approach avoids generating EGMF in cosmic voids.

In Ref.~\cite{dolag} constrained simulations of the local large
scale structure were performed and the magnetic smoothed particle
hydrodynamics technique was used to follow EGMF evolution. The
EGMF was seeded by a uniform seed field of maximal strength compatible
with observed rotation measures in galaxy clusters.

The questions considered in these two works were somewhat different,
however. In Ref.~\cite{dolag} deflections of UHECR above
$4\times10^{19}\,$eV were computed as a function of the direction
to their source which were assumed to be at cosmological distances.
This made sense, because (i) the constrained simulations gives a
viable model of our local cosmic neighborhood within about 100 Mpc,
at least on scales beyond a few Mpc and (ii) the deflections typically
were found to be smaller than a few degrees. Concrete source distributions
were not considered.

\begin{figure}[ht]
\includegraphics[width=0.9\textwidth,clip=true]{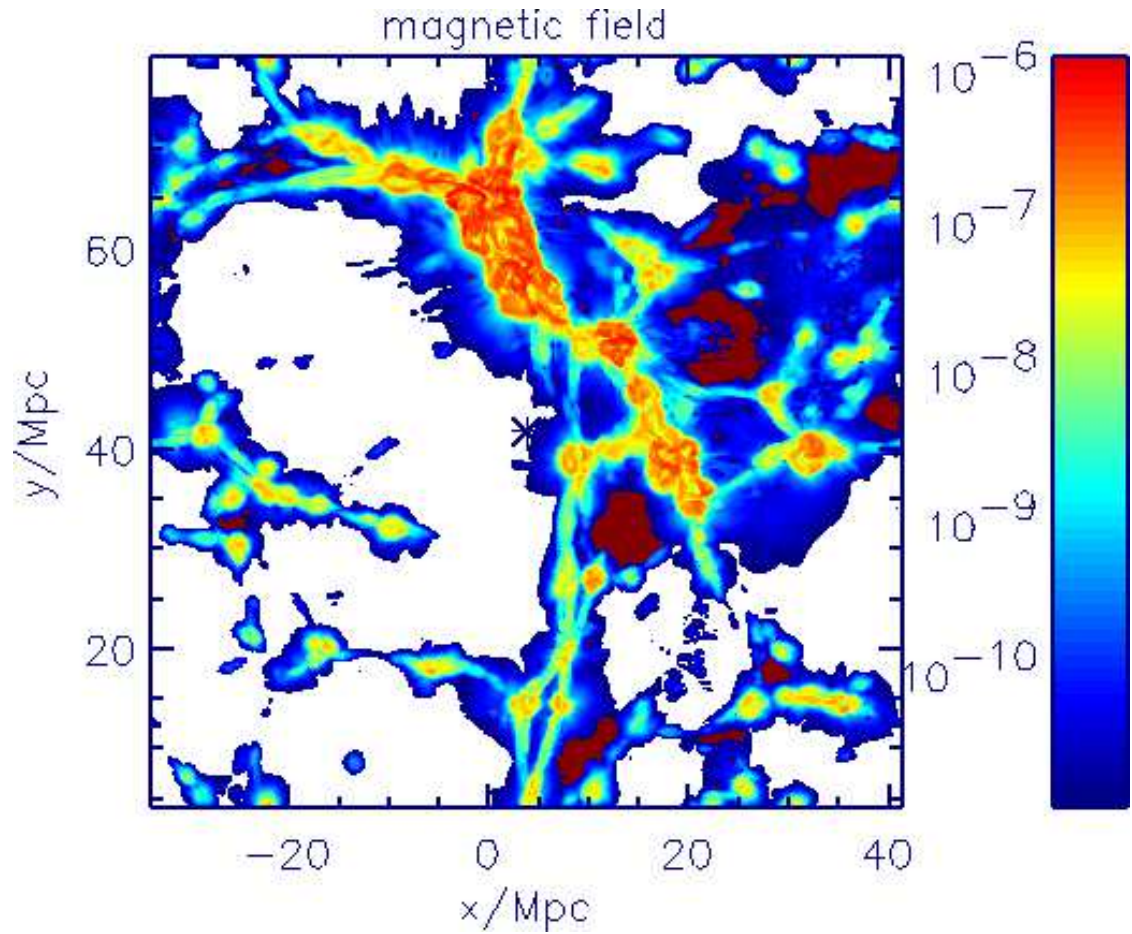}
\caption{Log-scale two-dimensional cut through
magnetic field total strength in Gauss (color
scale in Gauss) for a scenario
in good agreement with UHECR data studied in Ref.~\cite{sme}.
The observer is in the center of the figures and is marked by a star.
The EGMF strength at the observer is $\simeq10^{-11}\,$G.}
\label{fig10}
\end{figure}

In contrast, Ref.~\cite{sme} was not concerned with concrete sky
distributions or deflection maps because the simulation was unconstrained
and thus only gave a typical large scale structure model and not our
concrete local neighborhood. Instead, the question was asked which
observer positions and source distributions and characteristics
lead to UHECR distributions whose spherical multi-poles for $l\leq10$
and auto-correlation at angles $\theta\la20^\circ$ are consistent
with observations. As a result it was found that (i) the observed
large scale UHECR isotropy requires the neighborhood within a few Mpc
of the observer is characterized by weak magnetic fields below $0.1\,\mu$G,
and (ii) once that choice is made, current data do not strongly
discriminate between uniform and structured source distributions
and between negligible and considerable deflection. Nevertheless,
current data moderately favor a scenario in which (iii) UHECR
sources have a density $n_s\sim10^{-5}\,{\rm Mpc}^{-3}$ and follow the matter
distribution and (iv) magnetic fields are relatively pervasive within the large
scale structure, including filaments, and with a strength of order of a $\mu$G
in galaxy clusters. A two-dimensional cut through the
EGMF environment of the observer in a typical such scenario is
shown in Fig.~\ref{fig10}.

It was also studied in Ref.~\cite{sme} how future data of considerably
increased statistics can be used to learn more about EGMF and source
characteristics. In particular, low auto-correlations at
degree scales imply magnetized sources quite independent of
other source characteristics such as their density. The latter can
only be estimated from the auto-correlations halfway reliably
if magnetic fields have negligible impact on propagation.
This is because if sources are immersed
in considerable magnetic fields, their images are smeared out,
which also smears out the auto-correlation function over several
degrees. For a sufficiently high source density, individual images
can thus overlap and sensitivity to source density is consequently
lost. The statistics expected from next generation experiments
such as Pierre Auger~\cite{auger} and EUSO~\cite{euso} should
be sufficient to test source magnetization by the auto-correlation
function~\cite{sme}.

\begin{figure}[ht]
\includegraphics[width=0.9\textwidth,clip=true]{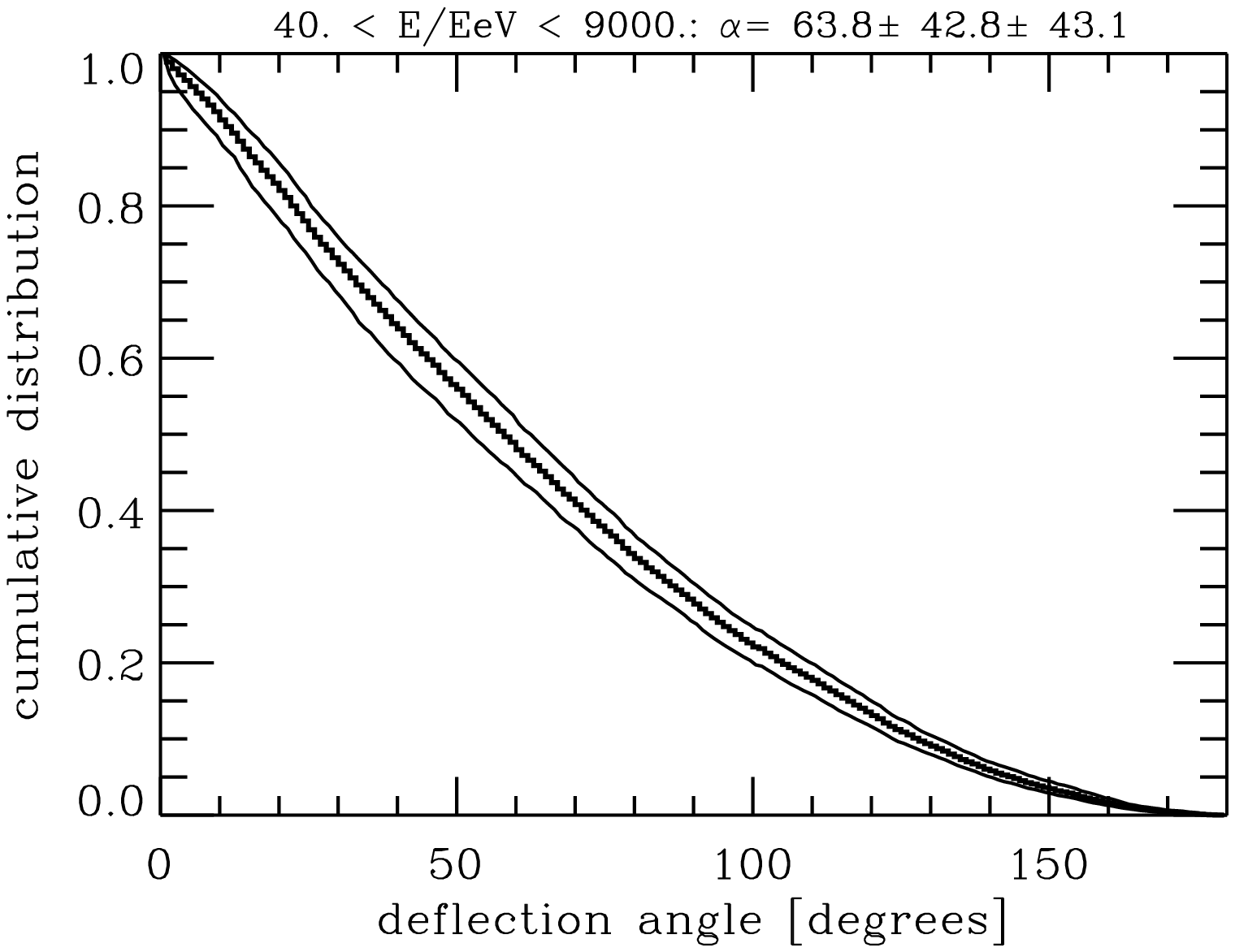}
\caption{The cumulative distribution of UHECR deflection
angles $\alpha$ with respect to the line of sight to the sources.
This is for a scenario from Ref.~\cite{sme} in good agreement with
UHECR data, where the sources follow the baryon density and have average
density $n_s=2.4\times10^{-5}\,{\rm Mpc}^{-3}$, and the EGMF included in
the large scale structure simulation reaches several micro Gauss in
the most prominent galaxy cluster. Shown are the average (middle,
histogram) and 1-$\sigma$ variations (upper and lower curves) above
$4\times10^{19}\,$eV, over 24 realizations varying in the positions
and luminosities $Q_i$ of individual sources,
the latter assumed to be distributed as $dn_s/dQ_i\propto Q_i^{-2.2}$
with $1\leq Q_i\leq100$ in arbitrary units. Also given on top
of the figure are average and variances of the distributions.}
\label{fig11}
\end{figure}

Interestingly, however, there is a considerable quantifiable difference
in the typical deflection angles predicted by the two EGMF scenarios
in Refs.~\cite{sme,dolag} that can {\it not} be compensated by
specific source distributions: Even for homogeneous source distributions,
the average deflection angle for UHECRs above $4\times10^{19}\,$eV
obtained in Ref.~\cite{sme} is much larger than in Ref.~\cite{dolag},
as can be seen in Fig.~\ref{fig11}. In fact, even if the magnetic field
strength is reduced by a factor 10 in the simulations of Ref.~\cite{sme},
the average deflection angle above $4\times10^{19}\,$eV is still
$\sim30^\circ$, only a factor $\simeq2.2$ smaller. This non-linear
behavior of deflection with field normalization is mostly due to the
strongly non-homogeneous character of the EGMF.

\begin{figure}[ht]
\includegraphics[width=0.9\textwidth,clip=true]{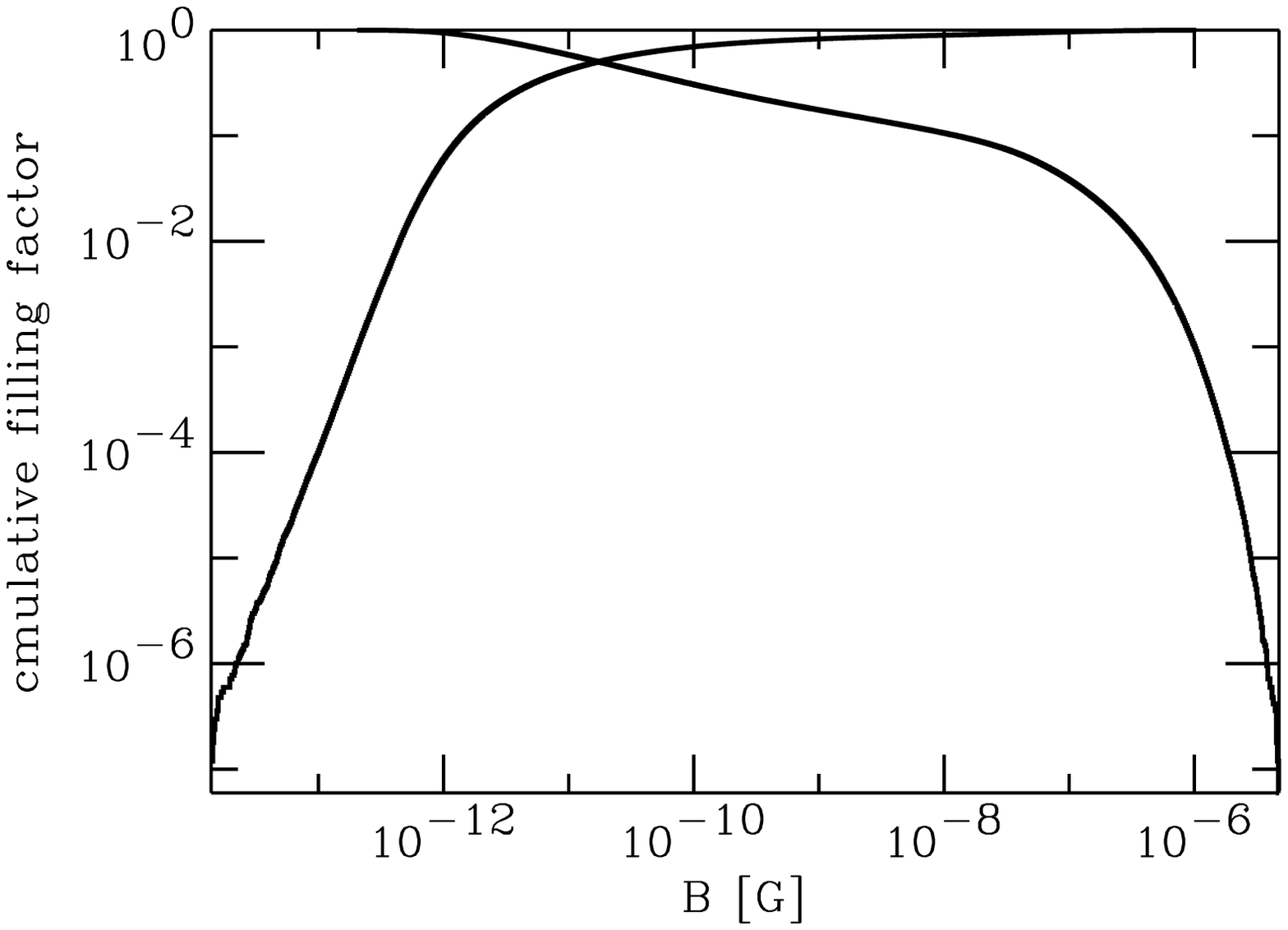}
\caption{The cumulative filling factors for EGMF strength
in the simulations used in Ref.~\cite{sme} above (decreasing curve)
and below (increasing curve) a given threshold, as a function of
that threshold.}
\label{fig12}
\end{figure}

Most of these differences are probably due to the different 
numerical models for the magnetic fields.
Although Ref.~\cite{dolag} start with uniform seed fields,
whereas in Ref.~\cite{sme} seed fields are injected
at shocks, by itself, this difference
should not influence the resulting EGMF very
much at late times, at least inside galaxy clusters~\cite{ryu}.
It should be noted, however, that in the filaments, where 
the gas motions are more uniform, the simulated magnetic fields may
depend to a certain extent on the initial seed fields although 
that is not trivial to quantify in general terms. 
In addition, numerical resolution
may play an important role because it affects the amplification and
the topological structure of the 
magnetic fields, both of which are important for the normalization
procedure, see below. The resolution
in Ref.~\cite{sme} is constant and much better in filaments and
voids but worse in the core of galaxy clusters than the (variable) 
resolution in Ref.~\cite{dolag}. If in both simulations the magnetic
fields are normalized to (or reproduce) the same ``observed'' values 
in the core of rich clusters then obviously their values 
in the filaments will be very different for the reasons outlined above.
This may partly explain why the contribution
of filaments to UHECR deflection is more important in Ref.~\cite{sme},
although a more detailed analysis and comparison are required to settle 
the issue. In any case, the magnetic fields obtained
in Ref.~\cite{sme} seem to be quite extended, as can be seen in
Fig.~\ref{fig12}: About 10\% of the volume is filled with fields
stronger than 10 nano Gauss, and a fraction of $10^{-3}$ is
filled by fields above a micro Gauss. Furthermore, typical deflection
angles change at most by a factor of 2 if magnetic field normalization
is decreased by a factor 10 or seed fields are chosen as uniform
in these simulations. The different amounts of
deflection obtained in the simulations of Refs.~\cite{sme,dolag} show
that the distribution of EGMF and their effects on UHECR propagation
are currently rather uncertain.

Finally we note that these studies should be extended to include
heavy nuclei~\cite{prepa} since there are indications that a fraction
as large as 80\% of iron nuclei may exist above $10^{19}\,$eV~\cite{watson}.
As a consequence, even in the EGMF scenario of Ref.~\cite{dolag}
deflections could be considerable and may not allow particle astronomy
along many lines of sight: The distribution of deflection angles in
Ref.~\cite{dolag} shows that deflections of protons above
$4\times10^{19}\,$eV of $\ga1^\circ$ cover a considerable fraction
of the sky. Suppression of deflection along typical lines of sight
by small filling factors of deflectors is thus unimportant in this
case. The deflection angle of any
nucleus at a given energy passing through such areas will therefore
be roughly proportional to its charge as long as energy loss
lengths are larger than a few tens of Mpc~\cite{bils}. Deflection angles of
$\sim20^\circ$ at $\sim4\times10^{19}\,$eV should thus be the rule
for iron nuclei. In
contrast to the contribution of our Galaxy to deflection which
can be of comparable size but may be corrected for within sufficiently
detailed models of the galactic field, the extra-galactic contribution
would be stochastic. Statistical methods are therefore likely to
be necessary to learn about UHECR source distributions and
characteristics. In addition, should a substantial heavy composition
be experimentally confirmed up to the highest energies, some sources would
have to be surprisingly nearby, within a few Mpc, otherwise only
low mass spallation products would survive propagation~\cite{er}.

The putative clustered component of the UHECR flux whose fraction
of the total flux seems to increase with energy~\cite{teshima1} may
play a key role in this context. It could be caused by discrete sources
in directions with small deflection. Spectrum and composition of the
flux from such sources could still by modified considerably by
magnetic fields concentrated around the source~\cite{sse,sigl}.
For example, since, apart from energy losses, cosmic rays
of same rigidity $Z/A$ are deflected similarly by cosmic magnetic
fields, one may expect that the composition of the clustered component
may become heavier with increasing energy. Indeed, in Ref.~\cite{teshima}
it was speculated that the AGASA clusters may be consistent with
consecutive He, Be-Mg, and Fe bumps.

It thus seems evident that the influence of large
scale cosmic magnetic fields on ultra-high energy cosmic ray propagation
is currently hard to quantify and may not allow to do ``particle
astronomy'' along most lines of sight, especially if a significant
heavy nucleus component is present above $10^{19}\,$eV.

\begin{theacknowledgments}
The material presented here is based on a university course on neutrino
physics taught by the author and on research work with various
collaborators of whom I would especially like to thank Torsten Ensslin,
Francesco Miniati and Dmitry Semikoz. Finally, I would like to thank
the organizers of the XIth Brazilian School of Cosmology and Gravitation
for a terrific school.
\end{theacknowledgments}

\newpage

\bibliographystyle{aipprocl}

\end{document}